\documentclass[3p,,preprint,12pt]{elsarticle}
\makeatletter\if@twocolumn\PassOptionsToPackage{switch}{lineno}\else\fi\makeatother

\usepackage{amsmath}
\usepackage{array}
\usepackage{marvosym}
\usepackage[ruled]{algorithm2e}
\usepackage{algpascal}
\usepackage{algc}
\usepackage{booktabs}
\usepackage{threeparttable}  
\usepackage{booktabs}
\usepackage{multirow}
\usepackage{caption}

\usepackage{color}
\usepackage{lineno}
\usepackage{colortbl,booktabs}

\usepackage{subfigure}
\usepackage{subcaption}
\usepackage{natbib}
\usepackage{amssymb}

\usepackage{tabulary,xcolor}
\usepackage{amsfonts}
\usepackage[T1]{fontenc}
\usepackage{hyperref}

\usepackage{comment}

\makeatletter
\let\save@ps@pprintTitle\ps@pprintTitle
\def\ps@pprintTitle{\save@ps@pprintTitle\gdef\@oddfoot{\footnotesize\itshape \null\hfill\thepage}}
\def\hlinewd#1{%
  \noalign{\ifnum0=`}\fi\hrule \@height #1%
  \futurelet\reserved@a\@xhline}

\AtBeginDocument{\ifNAT@numbers \biboptions{sort&compress}\fi}
\makeatother

\usepackage{ifluatex}
\ifluatex
\usepackage{fontspec}
\defaultfontfeatures{Ligatures=TeX}
\usepackage[]{unicode-math}
\unimathsetup{math-style=TeX}
\else 
\usepackage[utf8]{inputenc}
\fi 
\ifluatex\else\usepackage{stmaryrd}\fi

\usepackage{url,multirow,morefloats,floatflt,cancel,tfrupee}
\makeatletter

\AtBeginDocument{\@ifpackageloaded{textcomp}{}{\usepackage{textcomp}}}
\makeatother
\usepackage{colortbl}
\usepackage{xcolor}
\usepackage{pifont}
\usepackage[nointegrals]{wasysym}
\usepackage{floatrow}
\newtheorem{theorem}{Theorem}  
\newtheorem{lemma}{Lemma}

\newtheorem{definition}{Definition}
\makeatletter

\def\mcWidth#1{\csname TY@F#1\endcsname+\tabcolsep}

\def\cAlignHack{\rightskip\@flushglue\leftskip\@flushglue\parindent\z@\parfillskip\z@skip}
\def\rAlignHack{\rightskip\z@skip\leftskip\@flushglue \parindent\z@\parfillskip\z@skip}

\usepackage{ifxetex}
\ifxetex\else\if@twocolumn\usepackage{dblfloatfix}\fi\fi

\AtBeginDocument{
\expandafter\ifx\csname eqalign\endcsname\relax
\def\eqalign#1{\null\vcenter{\def\\{\cr}\openup\jot\m@th
  \ialign{\strut$\displaystyle{##}$\hfil&$\displaystyle{{}##}$\hfil
      \crcr#1\crcr}}\,}
\fi
}

\AtBeginDocument{%
  \@ifpackageloaded{endfloat}%
   {\renewcommand\efloat@iwrite[1]{\immediate\expandafter\protected@write\csname efloat@post#1\endcsname{}}}{\newif\ifefloat@tables}%
}%

\def\BreakURLText#1{\@tfor\brk@tempa:=#1\do{\brk@tempa\hskip0pt}}
\let\lt=<
\let\gt=>
\def\processVert{\ifmmode|\else\textbar\fi}

\@ifundefined{subparagraph}{
\def\subparagraph{\@startsection{paragraph}{5}{2\parindent}{0ex plus 0.1ex minus 0.1ex}%
{0ex}{\normalfont\small\itshape}}%
}{}

\newcommand\role[1]{\unskip}
\newcommand\aucollab[1]{\unskip}
  
\@ifundefined{tsGraphicsScaleX}{\gdef\tsGraphicsScaleX{1}}{}
\@ifundefined{tsGraphicsScaleY}{\gdef\tsGraphicsScaleY{.9}}{}
\def\checkGraphicsWidth{\ifdim\Gin@nat@width>\linewidth
	\tsGraphicsScaleX\linewidth\else\Gin@nat@width\fi}

\def\checkGraphicsHeight{\ifdim\Gin@nat@height>.9\textheight
	\tsGraphicsScaleY\textheight\else\Gin@nat@height\fi}

\def\fixFloatSize#1{}
\let\ts@includegraphics\includegraphics

\def\inlinegraphic[#1]#2{{\edef\@tempa{#1}\edef\baseline@shift{\ifx\@tempa\@empty0\else#1\fi}\edef\tempZ{\the\numexpr(\numexpr(\baseline@shift*\f@size/100))}\protect\raisebox{\tempZ pt}{\ts@includegraphics{#2}}}}

\AtBeginDocument{\def\includegraphics{\@ifnextchar[{\ts@includegraphics}{\ts@includegraphics[width=\checkGraphicsWidth,height=\checkGraphicsHeight,keepaspectratio]}}}

\DeclareMathAlphabet{\mathpzc}{OT1}{pzc}{m}{it}

\def\URL#1#2{\@ifundefined{href}{#2}{\href{#1}{#2}}}

\def\UrlOrds{\do\*\do\-\do\~\do\'\do\"\do\-}%
\g@addto@macro{\UrlBreaks}{\UrlOrds}

\edef\fntEncoding{\f@encoding}

\makeatother

\begin{document}

\begin{frontmatter}
	
\title{Prospect Theory Based Individual Irrationality Modelling and Behavior Inducement in Pandemic Control}   

    \author[1]{Wenxiang Dong}
    \author[1]{H. Vicky Zhao\corref{c-448b}}

    \cortext[c-448b]{Corresponding author. Email address: vzhao@tsinghua.edu.cn}
\address[1]{\textit{Department of Automation, BNRist, Tsinghua University, Beijing, 100084, China}}

\begin{abstract}

It is critical to understand and model the behavior of individuals in a pandemic, as well as identify effective ways to guide people's behavior in order to better control the epidemic spread. However, current research fails to account for the impact of users' irrationality in decision-making, which is a prevalent factor in real-life scenarios. Additionally, existing disease control methods rely on measures such as mandatory isolation and assume that individuals will fully comply with these policies, which may not be true in reality. Thus, it is critical to find effective ways to guide people's behavior during an epidemic.
To address these gaps, we propose a Prospect Theory-based theoretical framework to model individuals' decision-making process in an epidemic and analyze the impact of irrationality on the co-evolution of user behavior and the epidemic. Our analysis shows that irrationality can lead individuals to be more conservative when the risk of being infected is small, while irrationality tends to make users be more risk-seeking when the risk of being infected is high.
We then propose a behavior inducement algorithm to guide user behavior and control the spread of disease. 
Simulations and real user tests validate our proposed model and analysis, and simulation results show that our proposed behavior inducement algorithm can effectively guide users' behavior.
\end{abstract}

     
\end{frontmatter}
    
\section{Introduction}

The outbreak of COVID-19 has resulted in a severe public health crisis and significant economic losses. Governments worldwide have implemented various measures, including lockdowns and mandatory quarantines, to curtail the spread of the disease. However, individuals may not comply with these policies, as many have their own opinions and preferences. During such public health crises, individuals tend to act irrationally, such as excessive panic in the early stages of a disease outbreak or underestimation of the dangers of the disease later in its spread, which can significantly impact their decisions and ultimately affect the spread of the epidemic. Moreover, individuals' behavior and the pandemic mutually influence each other. For example, individuals' behavior such as wearing masks, social distancing, and isolating can impede the epidemic, while people tend to adopt protective behaviors when the pandemic is more severe. Therefore, it is crucial to model individuals' behavior during disease outbreaks and determine how to control the spread of the disease by guiding people's behavior without resorting to mandatory measures.

\subsection{Literature Review}
In the following, we will review recent works on epidemic control over networks, user behavior modeling during an epidemic, and irrational behavior modeling.
\subsubsection{Epidemic Control}
There are numerous works studying how to control the disease spread over networks, and many works attempt to limit the disease spread on a network by removing nodes. Wang et al. demonstrate that epidemic spread on a network is highly correlated with the largest eigenvalue of the graph's adjacency matrix \cite{wang2003epidemic}. As a result, existing works in \cite{tong2010vulnerability,2015Node2,2017Spectral,2018Group}  attempt to manipulate the adjacency matrix's eigenvalues by removing nodes to minimize the likelihood of disease outbreaks. Additionally, the works in \cite{birge2022controlling,wan2021multi,hota2021closed} explore macro-level approaches to control disease spread, such as restricting population movement or implementing proportional quarantine. However, these methods assume that individuals will always comply with the policies, which is often not the case in reality.
\subsubsection{Individual Behavior Modeling During an Epidemic}
There are prior works attempting to model individuals' behavior during an epidemic. The authors of \cite{zhang2014suppression, wu2012impact, bagnoli2007risk, shang2013modeling} believe that as the proportion of infected individuals in the environment increases, people are more likely to adopt protective behaviors. Zhang et al. assume that individuals would be more likely to take protective behavior when the proportion of infected neighbors is high \cite{zhang2014suppression},{and they analyze the effect of individual protective behavior on disease spread}. The proposed model in \cite{wu2012impact} assumes that an individual's adoption of protective behavior is influenced by the proportion of infected neighbors, as well as regional and global infection rates. The works in  \cite{funka2009spread, granell2013dynamical, granell2014competing} observe that information dissemination also affects individual protective behavior during the pandemic. However, these studies do not consider the common and critical issue of individual irrationality, which can significantly affect their decision-making process during an epidemic.
\subsubsection{Irrational Behavior Modeling}
Individuals often make irrational decisions when faced with risks, such as the risk of being infected during a pandemic. For example, many people may be overly panicked in the early stages of a pandemic and may underestimate the risk of the disease in the later stages. A challenging issue here is how to mathematically model such irrational behavior. Prospect theory offers theoretical models to quantify how individuals tend to overestimate small probabilities and underestimate high probabilities \cite{kahneman2013prospect, prelec1998probability, tversky1992advances, gonzalez1999shape, prelec2000compound}. It is crucial to analyze the impact of this irrationality on individuals' decisions. Studies in \cite{oraby2015bounded, hota2019game, 2020Perception} analyze the impact of irrationality on users' decisions on whether to get vaccination during an epidemic, and they assume that this is a one-time binary decision problem where users only make one binary decision during the entire epidemic. However, in reality, individuals have multiple protective behaviors available to them, such as wearing a mask, washing hands, and isolating at home, and they need to continuously decide whether or not to take such protective behaviors and which behavior to take during the entire epidemic.  
Individuals may choose to adopt the highest-level protective measures such as self-isolation when the epidemic is rapidly spreading, while they may decide not to take any protective behaviors when the epidemic is declining. Therefore, it is crucial to investigate how the epidemic and individual decisions continuously interact with and influence each other over time.

In summary, current studies on individual behavior modeling in epidemics either neglect the impact of irrational decision-making or fail to account for behavior changes in response to the epidemic. Moreover, existing research on epidemic control often assumes users' absolute compliance with the government's policies. Consequently, it is critical to consider user irrationality and the changes in individuals' decisions and propose an effective mechanism to guide their behaviors to control the epidemic.
\begin{figure}[t]
	\centering
	\begin{minipage}[t]{0.7\linewidth}
		\centering
		\includegraphics[width=1\linewidth]{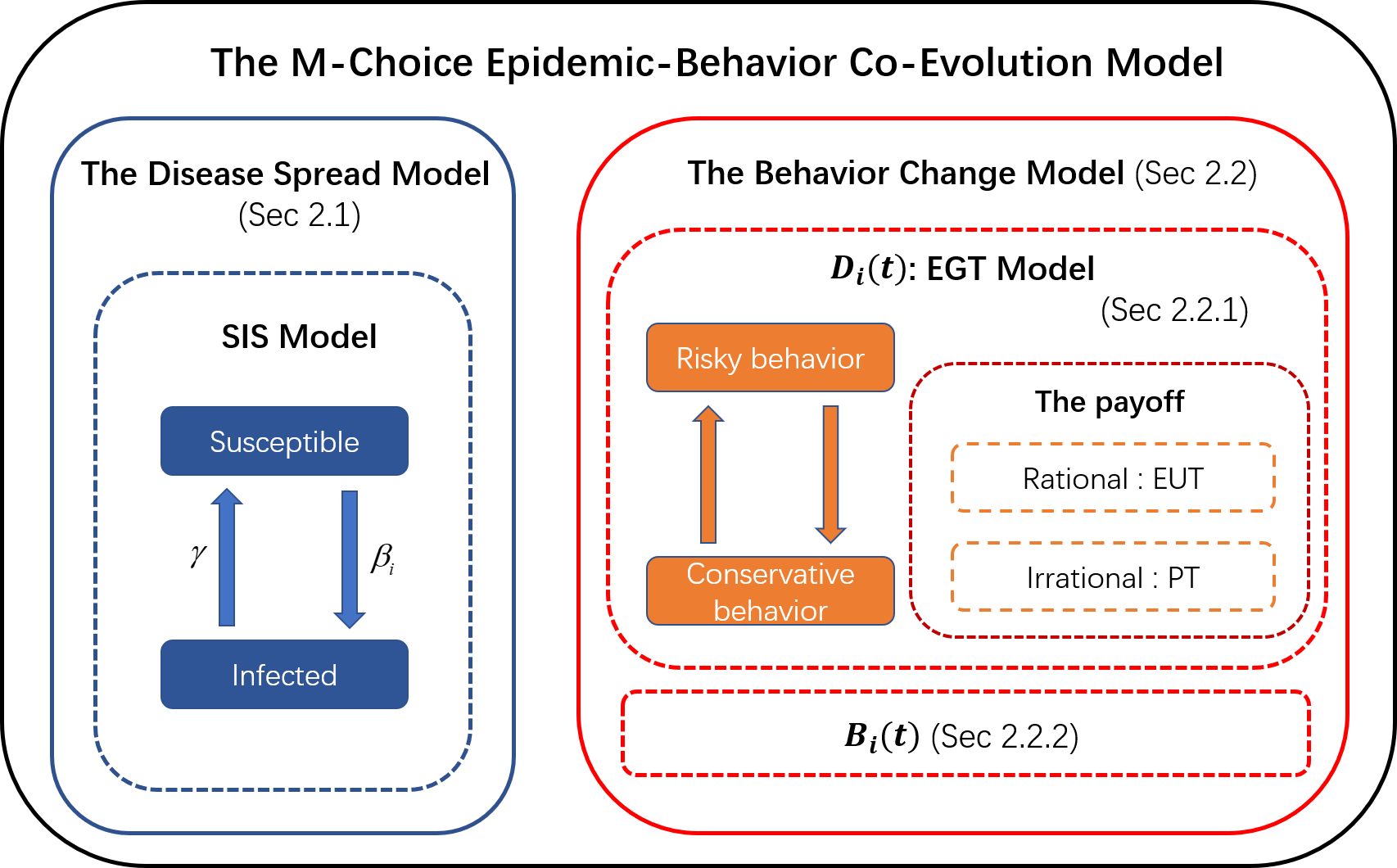}	
	\end{minipage}%
	\caption{The proposed M-choice epidemic-behavior co-evolution model.}
	\label{illustration}
\end{figure}
\subsection{Our Contribution}
In this paper, we model the individuals' irrational decisions during an epidemic and analyze the impact of the individuals' irrationality on their decisions as well as the epidemic. Based on the individuals' behavioral model, we propose an effective method to guide user behavior and control the spread of disease.

In summary, the main contributions of our work are:

\begin{itemize}
	\item We build an M-choice epidemic-behavior co-evolution model to simulate individuals' irrational decision-making and analyze their impact on the epidemic. We theoretically analyze the co-evolution of user behavior and epidemic and its steady state. We also study the impact of irrationality on individuals' behavior and disease spread.
	
	\item Given the above individual behavior model, we propose a behavior inducement algorithm to guide individuals' decisions to control the epidemic. 
	
	\item We validate our individual behavior model and behavior inducement algorithm through simulations. In addition, we use real user tests to validate the conclusions about the impact of irrationality on individuals' behavior. 
\end{itemize}

The rest of the paper is organized as follows. Section \ref{sec:Model}
presents our proposed epidemic-behavior co-evolution model. Section
\ref{sec:analysis} analyzes the steady state of the epidemic-behavior dynamics and the influence of irrationality.
Section \ref{sec:control} presents the behavior inducement method to control the disease spread.
Section \ref{sec:simulation} shows the simulation results.
Section \ref{sec:real} shows the results of real user tests, and conclusions are
drawn in Section \ref{sec:conclusion}.

\section{The M-Choice Epidemic-Behavior Co-Evolution Model}
\label{sec:Model}

During an epidemic, individual behavioral choices and the spread of the disease mutually influence each other. When the probability of infection and the potential losses are high, people tend to adopt protective behaviors. In turn, these protective behaviors can effectively inhibit the spread of the disease. In this section, we propose a model to capture the co-evolution of individual behavioral choices and disease spread during a pandemic. We consider irrational behavior in our model, and use the model with rational behavior assumption as a baseline to analyze the impact of irrationality on the evolutionary dynamics of the epidemic and its steady states. Building upon the model presented in \cite{poletti2009spontaneous}, we assume that individuals have a choice among $M$ possible behaviors based on the severity of the epidemic, and these behavioral choices, in turn, impact the spread of the disease.

Following the work in \cite{granell2013dynamical}, we use two undirected networks to represent the connections among individuals. The first network is the physical contact network where the disease spreads. The second network is the information network where individuals exchange information about their current health state and behavioral choices. It is worth noting that the two networks are different. In reality, an individual may get infected by strangers in a restaurant or on a bus, while their behavior will not be influenced by these strangers since they have not interacted with them. Similarly, users' decisions may be influenced by their friends on the Internet without any physical contact with each other. 
To simplify the analysis, we make the assumption that both the physical contact network and the information network are regular networks consisting of $N$ nodes, where each node represents an individual. In a regular network, each node has a fixed degree, denoted as $\bar{k}$ for the physical contact network and $\bar{d}$ for the information network. As individuals communicate with each other in the information network, we assume that they have knowledge of the health statuses of their neighbors in the information network. However, in the physical contact network, there is no exchange of information, and we assume that individuals do not have knowledge of the health statuses of their neighbors there. In the following, we will use the terms ``graph'' and ``network'' interchangeably, and the terms ``node'', ``user'', and ``individual'' interchangeably.

Our model consists of two interconnected parts: the disease spread model and the behavior change model. The disease spread model quantifies how the pandemic propagates through the network given the current behaviors of all individuals, and the behavior change model describes how individuals update their behaviors based on the current number of infected individuals and the behaviors of their neighbors. 
The illustration of our model is in Fig.\ref{illustration}. In the following, we will introduce these two components of our model in detail.


\subsection{The Disease Spread Model} \label{sec:SISmodel}
We use the classic Susceptible-Infected-Susceptible (SIS) model to depict the spread of the disease. Each individual can be in one of two health states: susceptible or infected. We divide time into slots of equal length. At each time slot, a susceptible individual can be infected by an infected individual at a given infection rate, and an infected individual recovers at a certain recovery rate. We assume that susceptible individuals can adopt several different protective measures to reduce their risk of infection, and assume that the susceptible individuals have a total of $M$ possible behavioral options $\{a_1, a_2, ..., a_M\}$. For example, some may take the pandemic very seriously and adopt self-quarantine to avoid contact with infected people; some may adopt medium-level precautions such as wearing masks when going out and washing hands frequently; while others may take no protection and act as if the epidemic does not exist. To simplify the analysis, we also assume that all susceptible persons taking action $a_i$ have the same infection rate $\beta_i$. For those already infected with the disease, we consider the worst-case scenario where they do not take any protective behavior such as home isolation to prevent the disease from further spreading. This assumption is grounded in the understanding that infected individuals might not possess the same level of motivation or necessity to embrace protective measures since they are already infected. We also assume that all infected people have the same recovery rate $\gamma$ to simplify the analysis.
Let $s(t)$ and $i(t)$ denote the fractions of susceptible and infected individuals, respectively, at time $t$, while $x_i(t)$ denotes the fraction of people adopting action $a_i$ among all susceptible individuals at time $t$. Then the mean-field equation of the disease spread is:
\begin{equation}
\begin{split}
\dfrac{ds}{dt} = &\gamma i(t)-s(t)i(t)\bar{\beta} \bar{k},
\\ \mbox{and} \quad \dfrac{di}{dt} = &s(t)i(t)\bar{\beta} \bar{k}-\gamma i(t),
\end{split} \label{eqn:SIS}
\end{equation}
where $s(t)+i(t)=1$ and $\overline{\beta}=\sum_{i=1}^n \beta_i x_i(t)$. Then, by plugging $s(t)=1-i(t)$ into (\ref{eqn:SIS}), we can get the differential equation modeling the disease spread
\begin{equation}
\dfrac{di(t)}{dt} = i(t)(1-i(t))\overline{\beta} \bar{k}-\gamma i(t).
\label{eq:disease}
\end{equation}

\subsection{The Behavior Change Model} \label{sec:behaviorchangemodel}

The individual behavior model quantifies the dynamics of $\{ x_i(t) \}$, which represents the proportion of susceptible individuals adopting action ${ a_i }$ at time $t$. The change in ${x_i(t)}$ can be attributed to two main factors. First, individuals may alter their decisions over time in response to the severity of the pandemic and the influence of their neighbors' behaviors. We use $D_i(t)$ to represent this part of the change in $x_i(t)$. In addition, due to nodes' changes in their health state, the proportion of susceptible individuals adopting different behaviors may also change. For example, a susceptible individual who was taking action $a_i$ at time $t-1$ may become infected at time $t$, or an infected person recovers at time $t$ and decides to take action $a_i$. We use $B_i(t)$ to represent this part of the change in $x_i(t)$. Then we have:

\begin{equation}
	\dfrac{dx_i(t)}{dt}=D_i(t)+B_i(t).
	\label{eq:dx}
\end{equation}

In Section \ref{EGT}, we focus on modeling $D_i(t)$, where individuals' decisions are influenced by their neighbors and the severity of the pandemic. In Section \ref{Spontaneous}, we study $B_i(t)$ and analyze how the changes in individuals' health states may affect the change in $x_i(t)$.


\subsubsection{Analysis of $D_i(t)$}
\label{EGT}
To model individuals' active behavior change in response to their neighbors' influence and the proportion of infected individuals, we employ evolutionary game theory. Evolutionary game theory is a useful framework to study the impact of neighbors on individuals' decisions \cite{jiang2014graphical,li2022graphical}. The basic elements of evolutionary game theory include individual, strategy, payoff, and strategy update rules. We will introduce these elements one by one in the following.

\noindent\textbf{Individual and Strategy:}
Each individual is represented as a node in the information network. As mentioned in Section \ref{sec:SISmodel}, we assume that there are a total of $M$ possible protective behaviors $\{a_1, \cdots, a_M\}$ for susceptible individuals, and each behavior $a_i$ corresponds to one strategy for a susceptible individual. For infected individuals, as mentioned in Section \ref{sec:SISmodel}, they do not take any protective behavior such as home isolation to prevent the disease from spreading. Therefore, in this work, we focus on the analysis of susceptible users' behavior and study how their decisions are influenced by their neighbors and the severity of the epidemic. In each time slot, $m$  percent of susceptible individuals are randomly chosen as focal individuals. These focal individuals observe and imitate their neighbors' behaviors. The remaining susceptible individuals maintain their actions unchanged during this time slot. 

\noindent\textbf{The Payoff:} 
In this paper, we study the protective behavior of susceptible individuals. Note that infected individuals have different health statuses from susceptible ones, and they use the same and fixed strategy of no protective behaviors. Therefore, in this work, we assume that susceptible individuals consider these infected users' decisions not valuable to them, and that the strategies of all susceptible individuals are solely influenced by their susceptible neighbors. So we define the payoff for susceptible individuals only, and ignore the payoffs of infected individuals, as they do not impact susceptible individuals' update of their strategies.

In each time slot, every susceptible individual receives a payoff determined by the chosen strategy and interactions with neighbors. In this paper, we consider two scenarios where the susceptible individual is rational and irrational. Therefore, we define the payoff of different behavior based on the Expected Utility Theory (EUT) \cite{simon1955behavioral} and Prospect Theory (PT) \cite{tversky1992advances}, respectively, where EUT models the individual as a rational person and PT considers the individual's irrationality. Here, following the prior work in \cite{2020Perception}, to simplify the analysis, we assume that either all individuals are rational or they are all irrational, and compare the two results to analyze the impact of users' irrationality on the co-evolution of individuals' behavior and the pandemic.


\emph{Rational individuals' payoff function:}
Expected Utility Theory (EUT) is an economic theory that models the decision-making of rational individuals. When an individual chooses a specific behavior, denoted as $a_i$, it leads to $L$ potential actual payoffs $o_{i,1}, o_{i,2}, ..., o_{i,L}$ with probabilities $p_{i,1}, p_{i,2}, ..., p_{i,L}$, respectively. It is worth noting that an individual may receive more than one actual payoff for their behavior, and $\sum_{j=1...L} p_{i,j} \neq 1$. For example, if a user decides to go out for dinner during an epidemic, they will receive a positive payoff from enjoying the fine cuisine, while they may also face a negative payoff if they become infected. In addition, from EUT, the perceived payoff may differ from the actual payoff. For example, the relationship between the perceived and actual payoffs is often not linear, and there is a phenomenon of diminishing marginal payoff \cite{kahneman2013prospect}. In prior works in EUT, the value function $u^{E}(x)$ is often used to model the relationship between the actual payoff $x$ and the perceived payoff $u^E(x)$. There have been various forms of $u^{E}(x)$ used in the previous works, including the simplest form of $u^{E}(x) = x$, as well as the power function and the exponential function form \cite{wakker2010prospect}. Thus in EUT, the payoff associated with choosing behavior $a_i$ is calculated as the expected utility, denoted as $U_i^{EUT}$ and is defined as:
\begin{equation}\label{eqn:EUTdef}
U_i^{EUT}=\sum_{j=1...L}u^{E}(o_{i,j})p_{i,j}.
\end{equation}
In our behavior modeling and epidemic control problem, every susceptible individual adopting protective behavior $a_i$ would get a fixed actual payoff of $c_i$, which is the payoff from the behavior itself, and the probability of getting this outcome is $1$. One example is the gain from enjoying the fine cuisine of dining outside during an epidemic. If the individual is infected at the next moment, it would get an additional loss of $c_n$ with $c_n<0$. For an individual adopting $a_i$, the probability to be infected at time $t$ is approximately $\bar{k}\beta_i  i(t)$ \cite{0DIRECTED}\footnote{We consider those epidemics with $\beta_i \ll 1$, such as SARS, MERS and common influenza \cite{wu2020nowcasting,zhu2012advanced,foster2021estimating}, and we assume that $\bar{k}\beta_i < 1$.}. Therefore, in our model, the individual who takes $a_i$ will get two potential actual payoffs. A payoff of $c_i$ with probability $1$, and a payoff of $c_n$ with probability $\bar{k}\beta_i  i(t)$. So the expected utility in (\ref{eqn:EUTdef}) becomes
\begin{equation}
U_i^{EUT}=u^{E}(c_i)+u^{E}(c_n)\bar{k}\beta_i  i(t).
\label{U_EUT}
\end{equation}

\emph{Irrational individuals' payoff function:}
Different from the Expected Utility Theory (EUT), the Prospect Theory (PT) takes into account the irrational tendencies exhibited by individuals when faced with uncertainty. In PT, individuals have a tendency to overestimate the probability of small risks and underestimate the probability of large risks \cite{kahneman2013prospect}. Therefore, not only do the actual and the perceived payoffs differ, but the actual and the perceived probabilities are also different when irrational users face uncertainties.

Similar to EUT, the value function $u^{P}(x)$ in PT can take different forms,
one commonly used form is the power function \cite{prelec2000compound} 
\begin{equation}
u^{P}(x)=
\left\{
\begin{aligned}
&x^\sigma,&& \mbox{if}\  x\ge 0,\\
&-\lambda(-x)^\sigma,&& \mbox{if}\  x< 0,\
\end{aligned}
\right.
\label{utilityfunc}
\end{equation}
where $\lambda$ reflects the individual's sensitivity to gain and loss, and $\sigma \in (0,1]$ reflects the curvature and shape of the value function. Our theoretical analysis in Section \ref{sec:analysis} and Section \ref{sec:control} do not depend on the specific form of $u^{P}(x)$.

In addition, instead of using the actual probability $p_i$, individuals' perceived probability is $\pmb{\omega}(p_i,\alpha)$, where $\pmb{\omega}(p,\alpha)$ is the probability weighting function.  
Following the prior work in \cite{prelec1998probability}, in this work, we use the following weighting function to describe the relationship between the perceived probability $\pmb{\omega}(p,\alpha)$ and the actual probability $p$:
\begin{equation} \label{eqn:probweightfunc}
\pmb{\omega}(p,\alpha)=e^{(-(-lnp)^\alpha)},\ \ p\in[0,1],\ \alpha \in (0,1],
\end{equation}
where $\alpha$ is the irrationality coefficient. A smaller $\alpha$ indicates that the individual is more irrational (or equivalently, less rational), and the difference between the actual and the perceived probabilities are larger. Note that when $\alpha =1$, $\pmb{\omega}(p,1) = p$ and the perceived and actual probabilities are the same. Also, $\pmb{\omega}(1,\alpha)=1$ from (\ref{eqn:probweightfunc}), and we define $\pmb{\omega}(0,\alpha)=0$. This ensures that $\pmb{\omega}(p,\alpha) \in [0,1]$ and $\pmb{\omega}(p,\alpha)$ is an increasing function of $p$. For simplicity, we use the mean-field method and assume all individuals have the same $\alpha$.



Given the probability weighting function in (\ref{eqn:probweightfunc}) and the value function $u^P(x)$, when an irrational individual chooses behavior $a_i$, which leads to $L$ different potential payoffs $\{ o_{i,j} \}$ with corresponding probabilities $\{ p_{i,j} \}$, respectively, the expected payoff is 
\begin{equation} 
U_i^{PT}=\sum_{j=1...L}u^{P}(o_{i,j})\pmb{\omega}(p_{i,j},\alpha).
\label{U_PT}
\end{equation}

In our problem, same as the analysis of $U_i^{EUT}$ in the above, the individual who chooses behavior $a_i$ will get two potential actual payoffs: a payoff of $c_i$ with probability $1$, and a payoff of $c_n$ with probability $\bar{k}\beta_i  i(t)$. So (\ref{U_PT}) becomes
\begin{equation}
U_i^{PT}=u^{P}(c_n)\cdot\pmb{\omega}[\bar{k}\beta_i  i(t),\alpha]+u^{P}(c_i).
\label{U_PT2}
\end{equation}

Note that if the value functions of EUT and PT are identical (i.e., $u^{E}(x)=u^{P}(x)$) and the irrationality coefficient $\alpha$ is set to 1, then PT degenerates to EUT.

\noindent\textbf{Strategy Update Rules:}
In each time unit, $mN_0(t)$ individuals are randomly selected as the focal individuals to update their strategies and others would keep their strategies unchanged, where $m$ is the fraction of individuals who are chosen as the focal individuals, and $N_0(t)$ is the total number of susceptible users at time $t$ in the network. The focal individuals tend to imitate their neighbors' behavior with a high payoff. Following the work in \cite{2004Coevolutionary}, given a focal individual $v$ with strategy $a_i$ and given $v$ randomly chooses a neighbor $w$ using strategy $a_j$, the probability the individual $v$ changes its strategy to $a_j$ is
\begin{equation} \label{eqn:probchange}
p(a_i\rightarrow a_j)=\frac{1}{2}+\frac{\omega}{2}\frac{1}{U_{max}}(U_j-U_i),
\end{equation}
where $\omega \in (0,1]$ measures the strength of selection, and $U_{max}$ is the normalization term to ensure $p(a_i\rightarrow a_j) \leq 1$. $U_i$ and $U_j$ are the payoffs of strategy $a_i$ and $a_j$, respectively.

In our work, we assume that individuals with different behaviors are uniformly distributed in the entire network. Therefore, the probability that the focal user $v$ adopts behavior $a_j$ is $x_j(t)$, which represents the proportion of susceptible individuals who choose behavior $a_j$ at time $t$ in the entire network. And the proportion of focal user $v$'s susceptible neighbors adopting behavior $a_i$ is the same as the proportion of susceptible users adopting behavior $a_i$ in the entire network. Then the probability that $x_i(t)$ increases by $\frac{1}{N_0(t)}$ due to individuals' strategy change is
\begin{equation}
p\left(\Delta x_i=\frac{1}{N_0(t)} \right)=\sum_{j=1}^M x_i(t)x_j(t)p(a_j\rightarrow a_i),
\end{equation}
where $N_0(t) = N_0 s(t)$ is the number of susceptible individuals at time $t$. Similarly, the probability that $x_i(t)$ decreases by $\frac{1}{N_0(t)}$ due to individuals' strategy change is
\begin{equation} \label{eqn:DiMinus}
p\left(\Delta x_i=-\frac{1}{N_0(t)} \right)=\sum_{j=1}^M x_i(t)x_j(t)p(a_i\rightarrow a_j).
\end{equation}

Combining (\ref{eqn:probchange}) and (\ref{eqn:DiMinus}), we have
\begin{equation}
\begin{split}
D_i(t)=&mN_0(t)\left\{ p \left(\Delta x_i=\frac{1}{N_0(t)} \right) \times \frac{1}{N_0(t)} - p\left(\Delta x_i=-\frac{1}{N_0(t)} \right) \times \frac{1}{N_0(t)}\right \}\\
=&\frac{m\omega}{U_{max}}\sum_{j=1}^M x_i(t)x_j(t) (U_i-U_j).
\label{eq:Ei}
\end{split}
\end{equation}

\subsubsection{Analysis of $B_i(t)$}
\label{Spontaneous}

In reality, even if individuals do not change their behaviors, the proportion of susceptible individuals with different behaviors will change over time due to transitions in health states. Let $s_i(t)$ be the fraction of individuals who are susceptible and adopt behavior $a_i$ at time $t$ among the entire population, that is, $s_i(t)=s(t)x_i(t)$, where $s(t)$ is the fraction of susceptible individuals among all users in the network, and $x_i(t)$ is the fraction of individuals adopting $a_i$ among all the susceptible individuals. Note that $s(t)=\sum_{j=1}^{M}s_j(t)$, when all individuals do not change their behavior, we have
\begin{equation}
\begin{split}
B_{i}(t)=\frac{d}{dt} \left ( \frac{s_{i}(t)}{s(t)}\right )
=\frac{d}{dt} \left ( \frac{s_{i}(t)}{\sum_{j=1}^{M}s_j(t)}\right )=\frac{\sum_{j=1}^{M}[s_{i}'(t)s_j(t)-s_{j}'(t)s_i(t)]}{s^2(t)}.
\end{split}
\label{eq:Nit}
\end{equation}

Note that $s_i'(t)$, the first order derivative of $s_i(t)$, contains two parts. The first part represents the change caused by the infection of susceptible individuals, while the second part represents the change caused by the recovery of infected individuals. For the first part, as there are a total of $s_i(t)$ susceptible individuals adopting behavior $a_i$, and each of them has probability $\beta_i \bar{k} i(t)$ to be infected, we have $s_{i1}'(t) = -s_i(t)\beta_i \bar{k} i(t)$. For the second part, we assume that the recovered individuals would choose their behaviors based on the ratio of different behaviors of susceptible individuals in the network, similar to the work in \cite{2020Decisions}. Therefore, $\gamma i(t)$ infected individuals will recover, $x_i(t)$ of whom will choose action $a_i$, and we have $s_{i2}'(t) = \gamma x_i(t)i(t)$. By combining these two parts ($s_i'(t) = s_{i1}'(t) + s_{i2}'(t)$), we have
\begin{equation}
 s_i'(t)=-s(t)x_i(t)\beta_i \bar{k} i(t)+\gamma x_i(t)i(t).
\label{eq:st}
\end{equation}
Given \eqref{eq:Nit}, \eqref{eq:st} and $s_i(t)=s(t)x_i(t)$, we have
\begin{equation}
B_i(t)=\sum_{j=1}^{M}x_i(t)x_j(t)\bar{k} i(t)(\beta_j-\beta_i).
\label{eq:Bi}
\end{equation}


\subsubsection{The Overall Behavior Change Dynamics}
Combining \eqref{eq:dx}, \eqref{eq:Ei} and \eqref{eq:Bi}, the complete differential equations describing the dynamics of individual behavior change are
\begin{align}
\dfrac{dx_i(t)}{dt} =&\sum_{j=1}^{M}x_i(t)x_j(t)\bar{k} i(t)(\beta_j-\beta_i)+\frac{m\omega}{U_{max}}\sum_{j=1}^M x_i(t)x_j(t)(U_i-U_j),\ i = 1...M.  \label{eq:behavior}
\end{align}

\subsection{The Dynamics and the Steady States of the Epidemic-Behavior Co-evolution Model}

Based on the disease spread equation \eqref{eq:disease} and the behavior change dynamics \eqref{eq:behavior}, we get the M-choice epidemic-behavior co-evolution model as follows:

\begin{align} 
\dfrac{di}{dt} =& i(t)(1-i(t))\overline{\beta} \bar{k}-\gamma i(t),\label{eq:M-model}\\ \mbox{and} \quad 
\dfrac{dx_i}{dt} =&\sum_{j=1}^{M}x_i(t)x_j(t)\bar{k} i(t)(\beta_j-\beta_i) +\frac{m\omega}{U_{max}}\sum_{j=1}^M x_i(t)x_j(t)(U_i-U_j),\ i = 1...M.\nonumber
\end{align}
Here, the first differential equation represents the dynamics of individuals' health states, and the subsequent $M$ equations model the changes in the proportions of susceptible individuals adopting each of the $M$ behaviors.

At the steady state, both the proportion of infected individuals $i(t)$ and the proportions of individuals adopting different behaviors $\{ x_i(t) \}$ reach a stable state where there are no further changes in $i(t)$ and $\{ x_i(t) \}$. Even if a small group of individuals becomes infected/recovered or changes their strategies, the steady state would be restored. We denote the steady state of the M-choice model as $(i^*,x_1^*, \cdots, x_{M-1}^*)$ with $x^*_{M}=1-\sum_{j=1}^{M-1}x^*_j$.

To find the steady state of our M-choice disease spread and behavior change model, we follow an approach that is similar to \cite{2020Decisions} and apply Lyapunov's first method \cite{lyapunov1992general}.

\begin{definition}
	The steady-state $(i^*,x_1^*, \cdots, x_{M-1}^*)$ satisfies: for $j=1, \cdots, M-1$,
	\begin{equation} \label{eqn:steadystateeq}
	\left.\dfrac{di}{dt}\right|_{i=i^*} = 0,\ \left.\dfrac{dx_j}{dt}\right|_{x_j=x_j^*} =0,\ Re(\lambda_k)<0, \ k=1,...,M 
	\end{equation}
\end{definition}
where $\{ \lambda_k \}$ are the eigenvalues of the Jacobian matrix 
\begin{equation}
\left.\begin{bmatrix}
\frac{\partial i'}{\partial i} & \frac{\partial i'}{\partial x_1} & \cdots & \frac{\partial i'}{\partial x_{M-1}} \\
\frac{\partial x_1'}{\partial i} & \frac{\partial x_1'}{\partial x_1} & \cdots & \frac{\partial x_1'}{\partial x_{M-1}} \\
\vdots  & \vdots  & \ddots & \vdots  \\
\frac{\partial x_{M-2}'}{\partial i} & \frac{\partial x_{M-2}'}{\partial x_1} & \cdots & \frac{\partial x_{M-2}'}{\partial x_{M-1}} \\
\frac{\partial x_{M-1}'}{\partial i} & \frac{\partial x_{M-1}'}{\partial x_1} & \cdots & \frac{\partial x_{M-1}'}{\partial x_{M-1}} \\
\end{bmatrix}\right|_{(i^*,x_1^*, \cdots, x_{M-1}^*)}, 
\end{equation}
and $Re(x)$ means the real part of $x$.
As it is usually difficult to find the closed-form solution of (\ref{eqn:steadystateeq}), we often use numerical methods to find the steady states of the co-evolution of disease spread and behavioral choice.

\section{Analysis of the Steady State and the Influence of Irrationality for the 2-Behavior Model}
\label{sec:analysis}

To obtain insights into the co-evolution process of disease spread and behavioral choice and their steady states, in this section, we consider a simple scenario where each susceptible individual can have two possible actions and the theoretical solution of (\ref{eqn:steadystateeq}) can be obtained. For example, a susceptible individual may either take risky behavior such as continuing to go out in spite of the epidemic, or take conservative behavior such as home isolation. Home isolation helps significantly reduce the risk of being infected while it also leads to substantial economic loss as well as impacts people's physical and mental well-being. During the pandemic, individuals need to choose between high-cost low-risk conservative behavior and low-cost high-risk risky behavior. Their decisions are often influenced by the severity of the epidemic and the potential loss due to home isolation. People tend to choose self-isolation when the pandemic poses a greater threat to their health; while they may be inclined to go out when the loss due to home isolation is too high (e.g., losing their jobs and income). We use this simple scenario to gain insights into the co-evolution process and its steady states, and theoretically analyze the impact of irrationality on the pandemic as well as users' behavior. For the more general scenario with more than 2 possible behavior choices, we use numerical solutions and simulation results to show the evolution process, and an example with three possible behavior choices is shown in \ref{appendix:3behavior}.

In this section, based on our model in the previous section, we analyze the evolution of the epidemic and the dynamics of individuals' choices between two behaviors: risky behavior (going out) represented by $a_1$, and conservative behavior (home isolation) represented by $a_2$. As an example, we assume that the infection rate for risky behavior is $\beta_1$, while the infection rate for conservative behavior is $\beta_2 = 0$ as isolation ensures no infection risk. Then we analyze the steady states when the individuals are all rational or irrational, respectively, and compare their results to investigate the influence of irrationality. 

\subsection{Steady State Analysis}
\subsubsection{The steady states with all rational individuals}
When all individuals are rational, the payoff is modeled by EUT in (\ref{U_EUT}). 
Since $x_1(t) + x_2(t)=1$, we can replace $x_2(t)$ by $1-x_1(t)$. With 2 possible behavior choices, and given the EUT payoff function (\ref{U_EUT}), the differential equations in  (\ref{eq:M-model}) becomes
\begin{equation}
\left\{
\begin{aligned}
&\dfrac{di}{dt}\ \ = \beta_1 \bar{k} i(t) x_1(t) (1-i(t))-\gamma i(t),\\
&\dfrac{dx_1}{dt}=-\beta_1\bar{k}  x_1(t)(1-x_1(t)) i(t)
+ k_0 x_1(t)(1-x_1(t)) \cdot(u^{E}(c_1)+u^{E}(c_n) \beta_1 \bar{k} i(t)-u^{E}(c_2)),
\end{aligned}
\right.
\label{eq:EUT}
\end{equation}
where $k_0=\frac{m\omega}{U_{max}}>0$.
To find the steady states of (\ref{eq:EUT}), we have the following Theorem \ref{th1}.

\begin{theorem}
	The steady state $(i^E,x_1^E)$ of (\ref{eq:EUT}) satisfies \eqref{stable1},
 \begin{small}
	\begin{equation}
	(i^E,x_1^E)=
	\left\{
	\begin{aligned}
	&(0,1),&& \mbox{if}\ \bar{k}<\frac{\gamma}{\beta_1},\\
	&\left(1-\frac{\gamma}{\bar{k}\beta_1},1\right),&& \mbox{if}\ \bar{k}>\frac{\gamma}{\beta_1},\Phi_1 < 0, \\
	&\left(i^{(1)},\frac{\gamma}{(1-i^{(1)})\bar{k}\beta_1}\right),&& \mbox{if}\ \bar{k}>\frac{\gamma}{\beta_1}, \Phi_1 \ge 0,
	\end{aligned}
	\right.
	\label{stable1}
	\end{equation}
 \end{small}
	\begin{align}
		&\mbox{where} \   i^{(1)}=\dfrac{k_0(u^{E}(c_2)-u^{E}(c_1))}{(k_0u^{E}(c_n)-1)\bar{k}\beta_1}, 
         \  \Phi_1=\ -k_0(u^{E}(c_1)-u^{E}(c_2))+(\gamma-\bar{k}\beta_1)(k_0u^{E}(c_n)-1). \nonumber \\
         &\mbox{When } \bar{k}=\frac{\gamma}{\beta_1}, \mbox{ there is no steady state.} \nonumber
	\end{align}
  
	\label{th1}
\end{theorem}

\emph{Proof}: See \ref{appendix:th1}.

From Theorem \ref{th1}, There are three possible steady states, which correspond to three different situations in reality:
\begin{itemize}
	\item Case 1: The steady state $(0,1)$ represents the extreme situation where the infection rate is too low and the disease would die out eventually even without any protection. So all individuals choose the risky behavior of going out. The evolution process reaches the steady state $(0,1)$ when $\bar{k}<\frac{\gamma}{\beta_1}$, which is equivalent to $\frac{\beta_1}{\gamma}<\frac{1}{\bar{k}}$.
	The term $\frac{1}{\bar{k}}$ is the epidemic threshold of a homogeneous network \cite{2007Epidemic}. If $\frac{\bar{\beta}}{\gamma}<\frac{1}{\bar{k}}$, the epidemic would die out; otherwise, it would spread out.
	Since $\bar{\beta}=\beta_1x_1(t) \in [0,\beta_1]$, in this scenario, the disease would die out no matter which behavior people choose.
	Therefore, all individuals would choose the risky behavior in this steady state since it gives a higher payoff. Therefore, when $\bar{k}<\frac{\gamma}{\beta_1}$, the stable state is $i=0$ and $x_1=1$. 
	
	\item Case 2: 	
	For the steady state $\left(1-\frac{\gamma}{\bar{k}\beta_1},1\right)$, there are two constraints that need to be satisfied. The first constraint is $\bar{k}>\frac{\gamma}{\beta_1}$, which means that if all individuals choose the risky behavior, the disease will spread out. The second constraint is $\Phi_1 < 0$. To understand the second constraint, note that when all individuals choose the risky behavior with $x_1=1$, the proportion of infected individuals will reach the stable state $\hat{i}=1-\frac{\gamma}{\bar{k}\beta_1}$, which represents the maximum extent to which the disease can spread (proof: see \ref{adx:A}). If for all possible values of $i$ in the range $[0, \hat{i}]$, we have $\frac{dx_1(t)}{dt}>0$ for all $x_1(t)\in (0,1)$\footnote{Note that from \eqref{eq:EUT}, if $x_1(t)= 0 \text{ or } 1$, we have $\frac{dx_1(t)}{dt}=0$.}, then more people will choose the risky behavior as time goes on, and ultimately all individuals will choose the risky behavior at the steady state with $x_1=1$. This may happen when the risky behavior's payoff is much higher than the conservative behavior's with $c_1 \gg c_2$, or when the cost of being infected $c_n$ is very low. Note that from (\ref{eq:EUT}), $\frac{dx_1}{dt}|_{0\leq i\leq \hat{i},0<x_1<1}>0$ is equivalent to $\Phi_1 <0$, where $\Phi_1$ is defined in \eqref{stable1}. Therefore, when $\bar{k}>\frac{\gamma}{\beta_1}$ and $\Phi_1<0$, the steady state $\left(1-\frac{\gamma}{\bar{k}\beta_1},1\right)$ is reached, where all peoples choose the risk behavior, and the proportion of infected people reaches the maximum level $\hat{i}$.

	\item Case 3: 
	The steady state $\left(i^{(1)},\frac{\gamma}{(1-i^{(1)})\bar{k}\beta_1}\right)$ represents the scenario other than the above two extreme cases. In this scenario, the disease does not extinct, nor does it spread to the maximum extent, and at the steady state, $0<i^{(1)}<\hat{i}$ of the individuals in the network will be infected. Meanwhile, $0<\frac{\gamma}{(1-i^{(1)})\bar{k}\beta_1}<1$ of susceptible individuals will choose the risky behavior. This happens when $\bar{k}>\frac{\gamma}{\beta_1}$ and $\Phi_1 \geq 0$.
	
\end{itemize}

\subsubsection{The steady states with irrational individuals}
Next, we consider the scenario where all individuals are ``irrational'', and model their payoff function using the Prospect Theory. Plugging in the PT utility function (\ref{U_PT2}) into \eqref{eq:behavior}, the dynamic of the epidemic and the behavior becomes:
\begin{small}

\begin{equation}
\left\{
\begin{aligned}
&\dfrac{di}{dt} = \beta_1 \bar{k} i(t)  x_1(t) (1-i(t))-\gamma i(t),\label{eq:PT}\\
&\dfrac{dx_1}{dt} =-\beta_1\bar{k}  x_1(t)(1-x_1(t)) i(t) 	+ k_0 x_1(t)(1-x_1(t))\cdot(u^P(c_1)+u^P(c_n) \cdot\pmb{\omega}[\beta_1 \bar{k} i(t),\alpha]-u^P(c_2)).
\end{aligned}
\right.
\end{equation}
\end{small}
Similarly, we can get the steady state of (\ref{eq:PT}) in Theorem \ref{th2}.
\begin{theorem}
	The steady state $(i^P,x_1^P)$ of (\ref{eq:PT}) satisfies:
	\begin{equation}
	(i^P,x_1^P)=
	\left\{
	\begin{aligned}
	&(0,1),&& \text{if}\ \bar{k}<\frac{\gamma}{\beta_1}, \\
	&\left(1-\frac{\gamma}{\bar{k}\beta_1},1\right),&& \text{if}\ \bar{k}>\frac{\gamma}{\beta_1},\Phi_2 < 0, \\
	&\left(i^{(2)},\frac{\gamma}{(1-i^{(2)})\bar{k}\beta_1}\right),&& \text{if}\ \bar{k}>\frac{\gamma}{\beta_1}, \Phi_2 \ge 0,
	\end{aligned}
	\right.
	\label{eq:steadyPT}
	\end{equation}
	\begin{align}
	& \mbox{where} \quad \Phi_2=\ -k_0(u^P(c_1)-u^P(c_2))-(\gamma-\bar{k}\beta_1)-k_0u^P(c_n)\cdot\pmb{\omega} [\bar{k}\beta_1 -\gamma,\alpha],\nonumber\\ &\mbox{and} \quad  i^{(2)} \; \text{satisfies} \quad 
	 k_0u^P(c_n)\cdot\pmb{\omega}[\bar{k}\beta_1i^{(2)},\alpha]-\bar{k}\beta_1i^{(2)}+k_0(u^P(c_1)-u^P(c_2))=0.\\
	 \label{eq:i2def}
  &\mbox{When } \bar{k}=\frac{\gamma}{\beta_1}, \mbox{ there is no steady state.} \nonumber
	\end{align}
	\label{th2}
\end{theorem}

\emph{Proof}: See \ref{appendix:th2}.

The three stable states in PT share similarities with those in EUT:
\begin{itemize}
	\item Case 1: The steady state (0,1) is the same as Case 1 under EUT. In this situation, the disease would always die out, then all individuals would choose the risky behavior.
	
	\item Case 2: The steady state $\left(1-\frac{\gamma}{\bar{k}\beta_1},1\right)$ is the same as Case 2 under EUT. In situations where the payoff for risky behavior is extremely high or when the loss of being infected is very small, all individuals would choose risky behavior, causing the disease to spread to its maximum extent. Note that the first constraints in (\ref{stable1}) and (\ref{eq:steadyPT}) are the same, while the second constraints are different as $\Phi_1 \neq \Phi_2$.
	
	\item Case 3: The steady state $\left(i^{(2)},\frac{\gamma}{(1-i^{(2)})\bar{k}\beta_1}\right)$ is similar to Case 3 under EUT, where the disease does not extinct, nor does it spread to the maximum range.
\end{itemize}

\subsection{Analysis of Individuals' Irrationality}
In this section, we analyze the influence of individuals' irrationality on the steady state. 
In the weighting function in (\ref{eqn:probweightfunc}), the irrationality coefficient quantifies the irrationality degree of individuals, and a smaller $\alpha$ indicates that individuals are more irrational with the difference between the actual and perceived risk being larger. To analyze the impact of the irrationality coefficient on people's behavior, we compare the steady states $(i^P, x_1^P)$ at different $\alpha$, and we have the following Theorem $\ref{th4}$.

\begin{theorem}
	Given the same set of system parameters (that is, $\beta_1$, $\gamma$, $k_0$, $c_1$, $c_2$, $c_n$ and $\bar{k}$) and the same value function $u^P(x)$, let $0 < \underline{\alpha}< \bar{\alpha} < 1$ be two irrationality coefficients, and $(\bar{i}^P,\bar{x}_1^P)$ and $(\underline{i}^P,\underline{x}_1^P)$ are the steady states with $\bar{\alpha}$ and $\underline{\alpha}$, respectively, where the individuals with $\bar{\alpha}$ have low irrationality and individuals with $\underline{\alpha}$ have high irrationality. Then we have:
	\begin{itemize}
		\item[3a.] When $\bar{k}<\frac{\gamma}{\beta_1}$, all individuals, regardless of their irrationality degree, will choose risky behavior with $\bar{x}_1^P = \underline{x}_1^P=1$, and the epidemic will eventually die out with $\bar{i}^P = \underline{i}^P=0$.

		\item[3b.] When $\bar{k} >\frac{\gamma}{\beta_1}$ and $\Phi_2$ is not greater than or equal to 0 simultaneously for $\bar{\alpha}$ and $\underline{\alpha}$.
  
  In addition, if $1 - \frac{\gamma}{\bar{k}\beta_1} \le \frac{1}{\bar{k}\beta_1e}$, there are two possibilities.
  \begin{itemize}
      \item When $\Phi_2<0$ for both $\bar{\alpha}$ and $\underline{\alpha}$, all individuals, regardless of their irrationality degree, will choose the risky behavior with $\bar{x}_1^P=\underline{x}_1^P=1$.

      \item When $\Phi_2<0$ for $\bar{\alpha}$ and $\Phi_2\ge 0$ for $\underline{\alpha}$, all individuals with low irrationality will choose the risky behavior with $\bar{x}_1^P=1$. On the other hand, only a subset of individuals with high irrationality will opt for risky behavior with $\underline{x}_1^P < 1$.

  \end{itemize}

        On the contrary, if $1 - \frac{\gamma}{\bar{k}\beta_1} \ge \frac{1}{\bar{k}\beta_1e}$, there are two possibilities.

          \begin{itemize}
      \item When $\Phi_2<0$ for both $\bar{\alpha}$ and $\underline{\alpha}$, all individuals, regardless of their irrationality degree, will choose the risky behavior with $\bar{x}_1^P=\underline{x}_1^P=1$.

      \item When $\Phi_2<0$ for $\underline{\alpha}$ and $\Phi_2\ge 0$ for $\bar{\alpha}$, all individuals with high irrationality will choose the risky behavior with $\underline{x}_1^P=1$. On the other hand, only a subset of individuals with low irrationality will opt for risky behavior with $\bar{x}_1^P < 1$.

  \end{itemize}

		\item[3c.] When $\bar{k} >\frac{\gamma}{\beta_1}$, $\Phi_2 \geq 0$ for both $\bar{\alpha}$ and $\underline{\alpha}$, 
        the epidemic neither dies out nor spreads to the maximum extent, and a fraction of individuals choose the risky behavior at the steady state. 
        \begin{itemize}
        \item In addition, if $\bar{i}^P\le\frac{1}{\bar{k}\beta_1e}$, we have $\bar{i}^P\ge \underline{i}^P,\bar{x}_1^P\ge \underline{x}_1^P$, that is, compared to individuals with high irrationality, fewer individuals with low irrationality would choose the conservative behavior. The proportion of individuals with low irrationality getting infected would be higher than that of those with high irrationality.

        \item On the contrary, if $\bar{i}^P\ge\frac{1}{\bar{k}\beta_1 e}$, we have $\bar{i}^P\le \underline{i}^P,\bar{x}_1^P\le \underline{x}_1^P$, that is, compared to individuals with high irrationality, more individuals with low irrationality would choose the conservative behavior. The proportion of individuals with low irrationality getting infected would be lower than that of those with high irrationality.
        \end{itemize}

	\end{itemize}
	\label{th4}
\end{theorem}

\emph{Proof}: See \ref{appendix:th4}.

To better understand Theorem \ref{th4}, note that in (\ref{eq:steadyPT}), Case 1 represents the situation where the infection rate is too low and the disease would eventually die out no matter how individuals choose their behavior. Therefore, individuals' irrationality will not affect the outcome when $\bar{k}<\frac{\gamma}{\beta_1}$, as stated in Theorem 3a.

For Theorem 3b, when $\bar{k} >\frac{\gamma}{\beta_1}$ and $\Phi_2$ is not greater than or equal to 0 simultaneously for $\bar{\alpha}$ and $\underline{\alpha}$, if $1 - \frac{\gamma}{\bar{k}\beta_1} \le \frac{1}{\bar{k}\beta_1e}$, the percentage of high irrationality individuals choosing the risky behavior will be less than or equal to the percentage of low irrationality individuals, that is, $\underline{x}_1^P \le \bar{x}_1^P$. This is because, in this scenario, the risk of being infected is low (i.e., $\bar{i}^P=\underline{i}^P=1 - \frac{\gamma}{\bar{k}\beta_1} \le \frac{1}{\bar{k}\beta_1e}$), and higher irrationality makes individuals overestimate this small probability of risk, causing them to be more conservative. On the contrary, if $1 - \frac{\gamma}{\bar{k}\beta_1} \ge \frac{1}{\bar{k}\beta_1e}$, the percentage of high irrationality individuals choosing the risky behavior will be larger than or equal to the percentage of low irrationality individuals, that is, $\underline{x}_1^P \ge \bar{x}_1^P$. This is because, in this scenario, the risk of being infected is high (i.e., $\bar{i}^P=\underline{i}^P=1 - \frac{\gamma}{\bar{k}\beta_1} \ge \frac{1}{\bar{k}\beta_1e}$), and higher irrationality makes individuals underestimate this large probability of risk, causing them to be more adventurous.

For Theorem 3c, when $\bar{k} >\frac{\gamma}{\beta_1}$, $\Phi_2 \geq 0$ for both $\bar{\alpha}$ and $\underline{\alpha}$, both $(\bar{i}^P, \bar{x}_1^P)$ and $(\underline{i}^P, \underline{x}_1^P)$ are in Case 3 in Theorem \ref{th2}, where some individuals get infected while the rest do not. In this case, if the risk of getting infected is low when stable (i.e., $\bar{i}^P\le \frac{1}{\bar{k}\beta_1e}$), higher irrationality can motivate individuals to adopt conservative behaviors as they tend to overestimate this small risk, resulting in a decrease in the probability of being infected. On the contrary, if the risk of getting infected is high when stable (i.e., $\bar{i}^P\ge \frac{1}{\bar{k}\beta_1e}$), higher irrationality can reduce individuals' cautiousness as they tend to underestimate this large risk, resulting in more people getting infected. 

Note that from Section \ref{sec:behaviorchangemodel}, if the value functions of EUT and PT are identical (i.e., $u^{E}(x)=u^{P}(x)$), then EUT can be considered as a special case of PT with $\alpha=1$. 
Therefore, we can also apply Theorem \ref{th4} to make comparisons between rational individuals (following EUT) and irrational individuals (following PT).

In summary, irrationality tends to make users become more extreme, that is, risk-averse when the risk is small and risk-seeking when the risk is high. 

\section{Behavior Inducement to Control the Disease Spread}
\label{sec:control}
In this section, based on our previous analysis in Section \ref{sec:Model} and \ref{sec:analysis}, we study how to guide users' behavior and control the spread of disease through policy design and develop effective behavior inducement algorithms. 

\subsection{The Optimal Behavior Inducement Algorithms} \label{sec:behaviorguidance}

We first discuss measures that governments can take to guide people's behavior during an epidemic. For example, they can incentivize or penalize certain behaviors, such as subsidizing risky behaviors (e.g., going out) to boost the economy, penalizing risky behavior, or encouraging conservative behaviors (such as staying at home and wearing masks) to control the disease spread. In our model, this means the parameters $c_1$ and $c_2$ can be changed. In addition, during a pandemic, people often have different perceptions of the loss of being infected, which are largely due to the various propaganda efforts. So we assume that the parameter $c_n$ can also be adjusted. Furthermore, note that propaganda via social networks and media often influences people's irrationality \cite{atwan2019digital}, and thus, we assume that the irrationality coefficient $\alpha$ can be changed as well. In this work, to simplify the analysis, we consider the simple scenario where these parameters $c_1$, $c_2$, $c_n$ and $\alpha$ can be changed to the desired values. We plan to study in our future work the more practical scenario where the optimization parameters are the actions government can take (such as rewarding or punishing specific behaviors through policies) instead of the exact values of these parameters.



Next, we discuss the goals of behavior guidance. The first goal is to control the spread of the disease at the steady state. For example, the government may wish to keep the number of infected people as low as possible. We represent the loss caused by the pandemic as $l_1(i^P)$, where $(i^P, x_1^P)$ is the steady state of PT. Also, if a large percentage of people take conservative behavior such as self-isolation, it will have a significant impact on the economy as well as people's mental health. Therefore, the second loss term we consider in our work is the loss due to such conservative behavior $l_2(x_1^P)$. Furthermore, note that changing the values of $c_1$, $c_2$, $c_n$, and $\alpha$ through behavior inducements such as propaganda, subsidies, and penalties will incur costs. In this work, the third goal is to minimize the cost associated with behavior guidance $l_3(\pmb{\delta})$, where $\pmb{\delta} =[\Delta {\alpha}, \Delta c_n, \Delta c_1, \Delta c_2]$ is the intervention vector quantifying the extent to which these variables are changed. Given $c_1$, $c_2$, $c_n$, and $\alpha$ before behavior guidance, the adjusted parameters are
\begin{equation}
\begin{split}
\alpha'=\alpha+\Delta \alpha,\ c_n'=c_n+\Delta c_n,\  c_1' = c_1+\Delta c_1,\  \mbox{and} \; c_2'=c_2+\Delta c_2.
\end{split}
\end{equation}

Since the irrationality coefficient should be in $(0,1]$ and the payoff of being infected should be negative, we have $0<\alpha + \Delta \alpha \leq 1$ and $c_n + \Delta c_n < 0$. Our goal is to find the optimal $\pmb{\delta}$ to minimize the total loss.
The optimization problem is:
\begin{equation}
\begin{aligned}
& \underset{\pmb{\delta}}{\mathbf{min}}\ &&l_1(i^P(\pmb{\delta}))+l_2(x_1^P(\pmb{\delta}))+ l_3( \pmb{\delta})\\
&s.t.&&0<\alpha + \Delta \alpha \leq 1,\\
& &&c_n + \Delta c_n < 0.
\end{aligned}
\label{eq:orioptim}
\end{equation}
In this work, we do not specify the specific forms of $l_1(i^P)$, $l_2(x_1^P)$, and $l_3(\pmb{\delta})$, while we assume that they are differentiable, i.e., $\frac{\partial l_1}{\partial i}$, $\frac{\partial l_2}{\partial x_1}$, and $\frac{\partial l_3}{\partial \pmb{\delta}}$ exist. Moreover, we assume that $l_3(\pmb{\delta})$ satisfies 
\begin{equation}
\begin{cases}
\frac{\partial l_3}{\partial \Delta \alpha}>0 & \mbox{if}\ \ \Delta \alpha>0,\cr
\frac{\partial l_3}{\partial \Delta \alpha}<0 & \mbox{if}\ \ \Delta \alpha<0,
\end{cases}
\label{40}
\end{equation}
and the same constraint also holds for $\frac{\partial l_3}{\partial \Delta c_n}$, $\frac{\partial l_3}{\partial \Delta c_1}$ and $\frac{\partial l_3}{\partial \Delta c_2}$. 
This implies that when compared to the case where no behavior guidance is taken with $\pmb{\delta}=\mathbf{0}$, increasing or decreasing any of the variables ($c_1$, $c_2$, $c_n$, and $\alpha$) will result in an increase in $l_3(\pmb{\delta})$, and $l_3(\pmb{\delta})$ has the minimum value at $\pmb{0}$ with no behavior guidance. 

From \eqref{eq:steadyPT}, there are three possible steady states. To solve the optimization problem \eqref{eq:orioptim}, we need to consider all three possible steady states and analyze the optimal solution for each, which would be very complicated. Then we introduce Theorem \ref{th5} to simplify this problem.

Given the system parameters and $\pmb{\delta}$, let $(i^0,x_1^0)$ and $(i^P,x_1^P)$ be the steady states without and with behavior inducement, respectively. Let $\pmb{\delta}^{(3)}$ be the optimal adjustment parameter when the steady state after adjustment $(i^P,x_1^P)$ is in Case 3 in Theorem \ref{th2}. That means
\begin{equation}
\begin{aligned}
&\pmb{\delta}^{(3)} =  &&\underset{\pmb{\delta}}{argmin}\ l_1(i^P(\pmb{\delta}))+l_2(x_1^P(\pmb{\delta}))+ l_3( \pmb{\delta})\\
&s.t.&&0<\alpha + \Delta \alpha \leq 1,\\
& &&c_n + \Delta c_n < 0, \\
& &&  \bar{k}>\frac{\gamma}{\beta_1}, \Phi_2 \ge 0,
\end{aligned}
\label{eq:delta3}
\end{equation}
where $\Phi_2$ is defined in (\ref{eq:steadyPT}).

From our analysis in \ref{appendix:th5}, we have the following Theorem \ref{th5}.
\begin{theorem}
	For the optimization problem \eqref{eq:orioptim}, if $l_3(\pmb{\delta})$ satisfies \eqref{40}, given $\pmb{\delta}^{(3)}$ defined in \eqref{eq:delta3}, let $(i^0,x_1^0)$ and $(i^P,x_1^P)$ be the steady states without and with behavior guidance, respectively. Then the optimal solution of \eqref{eq:orioptim} is either $\pmb{0}$ or $\pmb{\delta}^{(3)}$. Specifically,
	\begin{itemize}
		\item if $i^0 =0$ and $x_1^0 =1$, that is, the original steady state without behavior guidance is Case 1 in \eqref{eq:steadyPT}, the optimal solution is $\pmb{0}$. That means no behavior guidance is needed, and the objective function in \eqref{eq:orioptim} is minimized at $\pmb{0}$.
		
		\item If $i^0 = 1-\frac{\gamma}{\bar{k}\beta_1}$ and $x_1^0 =1$, that is, the original steady state is Case 2 in \eqref{eq:steadyPT}, the optimal solution is either $\pmb{0}$ or $\pmb{\delta}^{(3)}$, and we can find the optimal solution by comparing $l_1(i^0) + l_2(x_1^0)$ with $l_1(i^P) + l_2(x_1^P) + l_3(\pmb{\delta}^{(3)})$.
  
		
		\item If $0<i^0<\hat{i}$ and $0<x_1^0<1$, that is, the original steady state is Case 3 in \eqref{eq:steadyPT}, the optimal solution is $\pmb{\delta}^{(3)}$.
	\end{itemize}
	\label{th5}
\end{theorem}

From Theorem \ref{th5}, we only need to model and solve the problem for Case 3 in (\ref{eq:steadyPT}), which greatly reduces the complexity of our problem. 

\subsection{Solving the Optimization Problems}

In this section, we consider the following scenario and use it as an example to demonstrate how to model and solve the optimization problem in \eqref{eq:orioptim}. Consider the scenario where the government wants to control the epidemic spread and reduce the impact on the economy and people’s mental health, that means, at the steady state after behavior inducement $(i^P, x_1^P)$, the percentage of infected people $i^P$ is no more than $i_m$ and the ratio of people taking risky behavior $x_1^P$ is at least $x_m$. That is, there are two constraints $0\leq i^P \le i_m$ and $1 \geq x_1^P \ge x_m$. Here, we assume that $0 \leq i_m \leq 1$ and $0 \leq x_m \leq 1$.

As we assume that the infection rate $\beta$, the recovery rate $\gamma$, and the average degree of the networks $\bar{k}$ are fixed and cannot be changed, it is possible that we cannot find $\pmb{\delta}$ that can make the steady-state $(i^P, x_1^P)$ satisfy both constraints simultaneously. For example, from the analysis in Section \ref{sec:analysis}, when the infection rate is very high, it is unlikely to control the epidemic to a very small range while everyone goes out without any protective measures. Therefore, given $\beta_1$, $\gamma$ and $\bar{k}$, the first step is to determine if the two constraints $0\leq i^P \le i_m$ and $1 \geq x_1^P \ge x_m$ are \emph{feasible}, that is, if it is possible to find $\pmb{\delta}$ that make the steady state satisfy both constraints. Section \ref{sec:feasibilitytest} studies how to determine the feasibility of the two constraints. If they are feasible, Section \ref{sec:boundedconstraints} explains the details of the optimization problem and proposes a fast algorithm to solve it. If they are not feasible, Section \ref{subsec:general algor} studies how to reformulate the problem and let the steady state $(i^P, x_1^P)$ be as close as possible to the constraints.


\subsubsection{The Feasibility Test} \label{sec:feasibilitytest}


From Theorem \ref{th5}, we only need to get the optimal solution in Case 3 and then we can get the optimal solution of the whole space by comparing it with $\pmb{0}$. From \eqref{eq:steadyPT}, if the steady state $(i^P, x_1^P)$ is in Case 3, it should satisfy 
\begin{equation}
\begin{aligned}
x_1^P = \frac{\gamma}{(1 - i^P)\bar{k} \beta_1}.
\label{eq:i_x}
\end{aligned}
\end{equation}
Note that in (\ref{eq:i_x}), $x_1^P$ is an increasing function of $i^P$. Therefore, given $0 \le i^P \le i_m$, we have $x_1^P = \frac{\gamma}{(1 - i^P)\bar{k} \beta_1} \in  [\frac{\gamma}{\bar{k} \beta_1},\frac{\gamma}{(1 - i_m)\bar{k} \beta_1}]$. If $x_m \leq \frac{\gamma}{(1 - i_m)\bar{k} \beta_1}$, then it is possible to find $\pmb{\delta}$ whose corresponding steady-state $(i^P, x_1^P)$ satisfies $0 \leq i^P \leq i_m$ and $1\ge x_1^P \geq x_m$ simultaneously, and thus, the two constraints are feasible. Otherwise, the two constraints are infeasible and cannot be satisfied at the same time.

\begin{figure}[t]
	
	\centering
	\subfigure{
 \scriptsize{(a)}
		\begin{minipage}[t]{0.5\linewidth}
			\centering
			\includegraphics[width=0.85\linewidth]{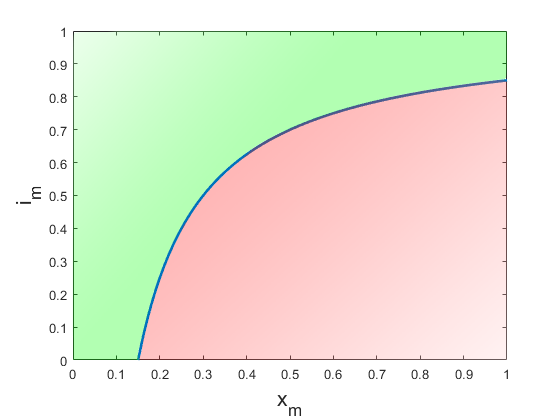}
			\label{i_x1}
		\end{minipage}%
	}%
	\subfigure{
 \scriptsize{(b)}
		\begin{minipage}[t]{0.5\linewidth}
			\centering
			\includegraphics[width=0.85\linewidth]{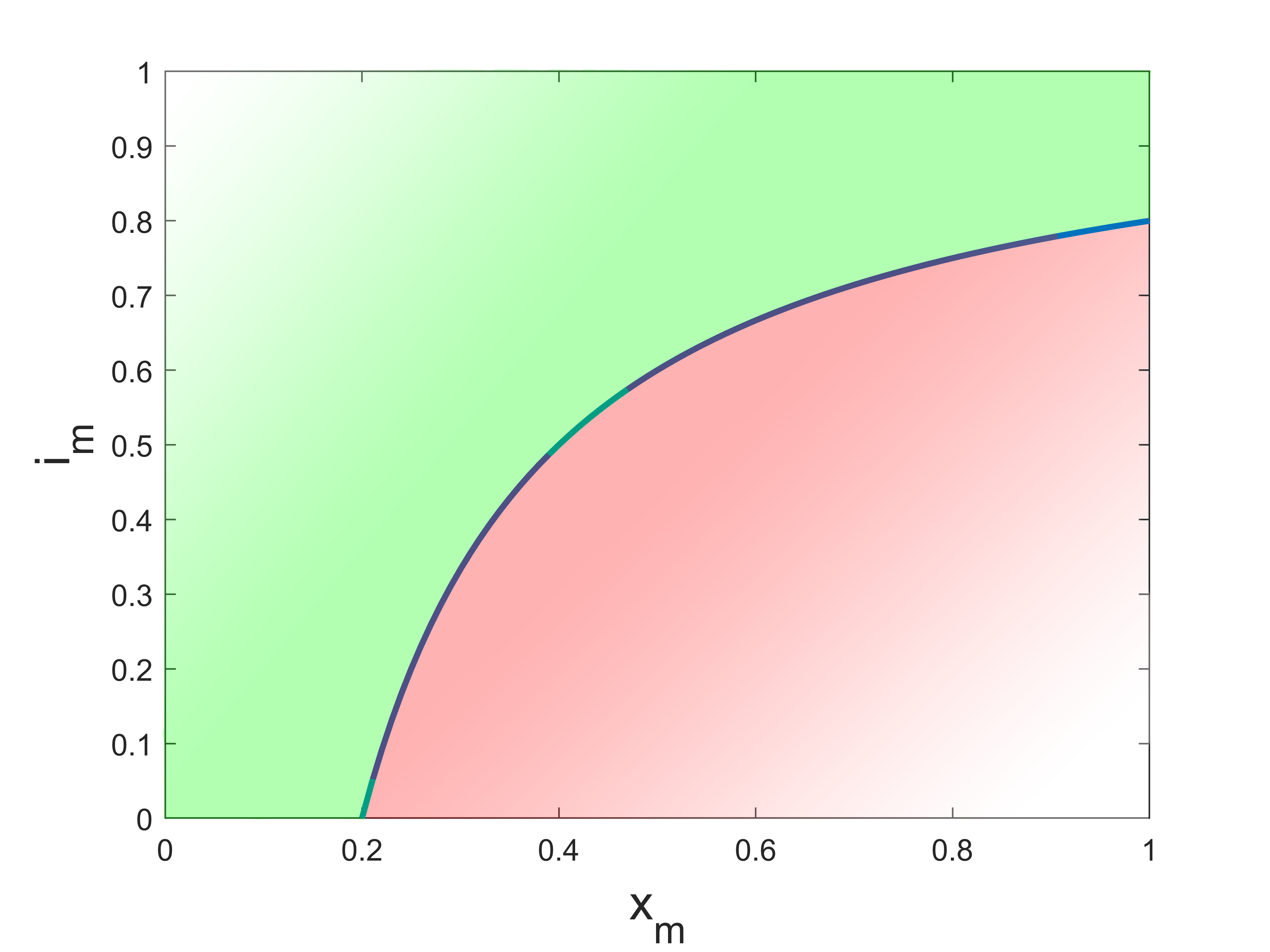}
			\label{i_x2}
		\end{minipage}%
	}%
	
	\centering
	\caption{The range of $i_m$ and $x_m$ when (a) $\frac{\beta_1}{\gamma}=\frac{2}{3}$ (b) $\frac{\beta_1}{\gamma}=\frac{1}{2}$ with $\bar{k}$ = 10. The green area represents the feasible constraints, and the red area represents the infeasible constraints.
	}
	\label{i_x}
\end{figure}

Fig. \ref{i_x} shows two examples of the feasible region of ($i_m$, $x_m$), where the green area includes all ($i_m$, $x_m$) where the constraints are feasible, and the red area includes all the infeasible constraints ($0\leq i^P \leq i_m$, $1\geq x_1^P \geq x_m$).

\subsubsection{Solving the Optimization Problem With Feasible Constraints} \label{sec:boundedconstraints}
If $0\leq i^P \le i_m$ and $1 \geq x_1^P \ge x_m$ are feasible, we discuss how to solve the optimization problem. Given the two constraints, we adopt the exterior-point method and transform the two constraints into a penalty function: 
\begin{equation}
    \mu  [P(i^P-i_m)+P(-i^P)+ P(x_m-x_1^P)+P(x_1^P-1)],
\end{equation}
where $\mu$ is a parameter that determines the intensity of the penalty, and $P(x){\buildrel \triangle \over =}(\max\{0, x\})^2$. We let $l_1(i^P)=\mu[ P(i^P-i_m)+P(-i^P)]$ and $l_2(x_1^P)=\mu[ P(x_m-x_1^P)+P(x_1^P-1)]$, then the optimization problem \eqref{eq:orioptim} becomes: 
\begin{equation}
\begin{aligned}
&&\underset{\pmb{\delta}}{\mathbf{\min}}\ &\mu  [P(i^P-i_m)+P(-i^P)+ P(x_m-x_1^P)+P(x_1^P-1)]+l_3( \pmb{\delta}),\\
&&s.t.\ \ & 0<\alpha + \Delta \alpha < 1,&&&\text{\ding{172}}\\
&& & c_n + \Delta c_n < 0.&&&\text{\ding{173}}\\
\end{aligned}
\label{bound1}
\end{equation}

According to Theorem \ref{th5}, the optimization problem can be transformed into the problem in Case 3 and then we can get the optimal solution of the whole space by comparing it with $\pmb{0}$. Here we only consider the situation where $\bar{k}>\frac{\gamma}{\beta_1}$.\footnote{If $\bar{k}<\frac{\gamma}{\beta_1}$, the optimal solution is $\pmb{0}$, and no calculation is required. So we do not consider it in our work} Then the optimization problem in Case 3 becomes:

\begin{equation}
\begin{aligned}
&&\underset{\pmb{\delta}}{\mathbf{min}}\ &\mu    [P(i^P-i_m)+P(-i^P)+ P(x_m-x_1^P)+P(x_1^P-1)]+l_3( \pmb{\delta}),\\
&&s.t.\ \ & 0<\alpha + \Delta \alpha < 1,&&&\text{\ding{172}}\\
&& & c_n + \Delta c_n < 0,&&&\text{\ding{173}}\\
&& &(1-i^P)\bar{k}\beta_1 x_1^P - \gamma = 0,&&&\text{\ding{174}}\\
&& &k_0(u^P(c_1+\Delta c_1)-u^P(c_2+\Delta c_2))+(\gamma-\bar{k}\beta_1)\\
&& & \qquad +k_0u^P(c_n+\Delta c_n)\cdot\pmb{\omega} [\bar{k}\beta_1 -\gamma,\alpha+\Delta \alpha] \leq 0,&&&\text{\ding{175}}\\
&& &k_0u^P(c_n+\Delta c_n)\cdot\pmb{\omega}[\bar{k}\beta_1 i^P,\alpha+\Delta \alpha]-\bar{k}\beta_1 i^P+k_0(u^P(c_1+\Delta c_1)-u^P(c_2+\Delta c_2))=0.&&&\text{\ding{176}}\\
\end{aligned}
\label{eq:bound}
\end{equation}
Here, $\pmb{\omega}[p,\alpha]=e^{(-(-lnp)^\alpha)}$. Constraints \ding{174}, \ding{175} and \ding{176} guarantees that the steady state is in Case 3 in Theorem \ref{th2}. The method for solving the problem is in \ref{appendix:dL}.

\subsubsection{Reformulation of the Optimization Problem When the Constraints are Infeasible} 
\label{subsec:general algor}
If the constraints $0\le i^P \leq i_m$ and $1\ge x_1^P \geq x_m$ are not feasible, we reformulate the optimization problem and make $(i^P$,$x_1^P)$ as close to $(i_m$,$x_m)$ as possible. Specifically, we transform the two constraints $0\le i^P \leq i_m$ and $1\ge x_1^P \geq x_m$ into $(i^P-i_m)^2+(x_1^P-x_m)^2$ in the objective function, so that the steady-state $(i^P, x_1^P)$ is close to $(i_m, x_m)$ in the $i-x_1$ plane. Then the optimization problem \eqref{eq:bound} becomes

\begin{equation}
\begin{aligned}
&&\underset{\pmb{\delta}}{\mathbf{min}}\ &(i^P-i_m)^2+(x_1^P-x_m)^2+l_3( \pmb{\delta}),\\
&&s.t.\ \ & 0<\alpha + \Delta \alpha < 1,\\
&& & c_n + \Delta c_n < 0,\\
&& &(1-i^P)\bar{k}\beta_1 x_1^P - \gamma = 0,\\
&& &k_0(u^P(c_1+\Delta c_1)-u^P(c_2+\Delta c_2))+(\gamma-\bar{k}\beta_1)\\
&& & \qquad +k_0u^P(c_n+\Delta c_n)\cdot\pmb{\omega} [\bar{k}\beta_1 -\gamma,\alpha+\Delta \alpha] \leq 0,\\
&& &k_0u^P(c_n+\Delta c_n)\cdot\pmb{\omega}[\bar{k}\beta_1 i^P,\alpha+\Delta \alpha]-\bar{k}\beta_1 i^P+k_0(u^P(c_1+\Delta c_1)-u^P(c_2+\Delta c_2))=0.\\
\end{aligned}
\label{eq:unbound}
\end{equation}

We can use the same method in Section \ref{sec:boundedconstraints} to solve \eqref{eq:unbound}, and details are shown in \ref{appendix:unbound}.


\section{Simulation Results}
\label{sec:simulation}
In this section, we first run simulations to validate our steady state analysis of the co-evolution process and the effect of irrationality on behavior and the pandemic in Section \ref{sec:analysis}. Then we validate the effectiveness of our behavior guidance algorithms proposed in Section \ref{sec:control}. As there are few previous works that analyze how to model irrational individuals in a pandemic, we do not compare our method with other works in this section.

\subsection{Simulations of the Steady States of EUT and PT}
Theorem \ref{th1} and \ref{th2} give theoretical analyses of the steady states when individuals are rational and irrational, respectively. To validate the two theorems, as an example, we conducted simulations on regular networks with 500 nodes. The physical contact network has a fixed degree of 10, while the information network has a degree of 20. We observe similar trends on other types of networks and with other parameters. We set the recovery rate to $\gamma=0.03$ as an example and let the infection rate $\beta_1$ vary. We first run the simulation to validate the three cases of the steady state of EUT and PT. Since risky behaviors such as going out and not wearing a mask are the default behaviors most people take in their daily lives, we set $c_1=0$; while conservative behaviors such as isolation and wearing masks can be regarded as behaviors with losses and we let $c_2<0$. In this work, we let $c_1=0$, $c_2=-1$, and $c_n=-20$. In order to facilitate the comparison between EUT and PT, we set $u^E(x)=u^P(x)$ and use the power function in \eqref{utilityfunc} with $\sigma=0.65$ and $\lambda=1$ as an example. For other value functions, we observe the same trend and omit the results here. For each simulation setup, we repeat the experiment 50 times and show the average result below.

\begin{figure}[h]
	
	\centering
	\subfigure[rational: $i^E$]{
		\begin{minipage}[t]{0.5\linewidth}
			\centering
			\includegraphics[width=0.85\linewidth]{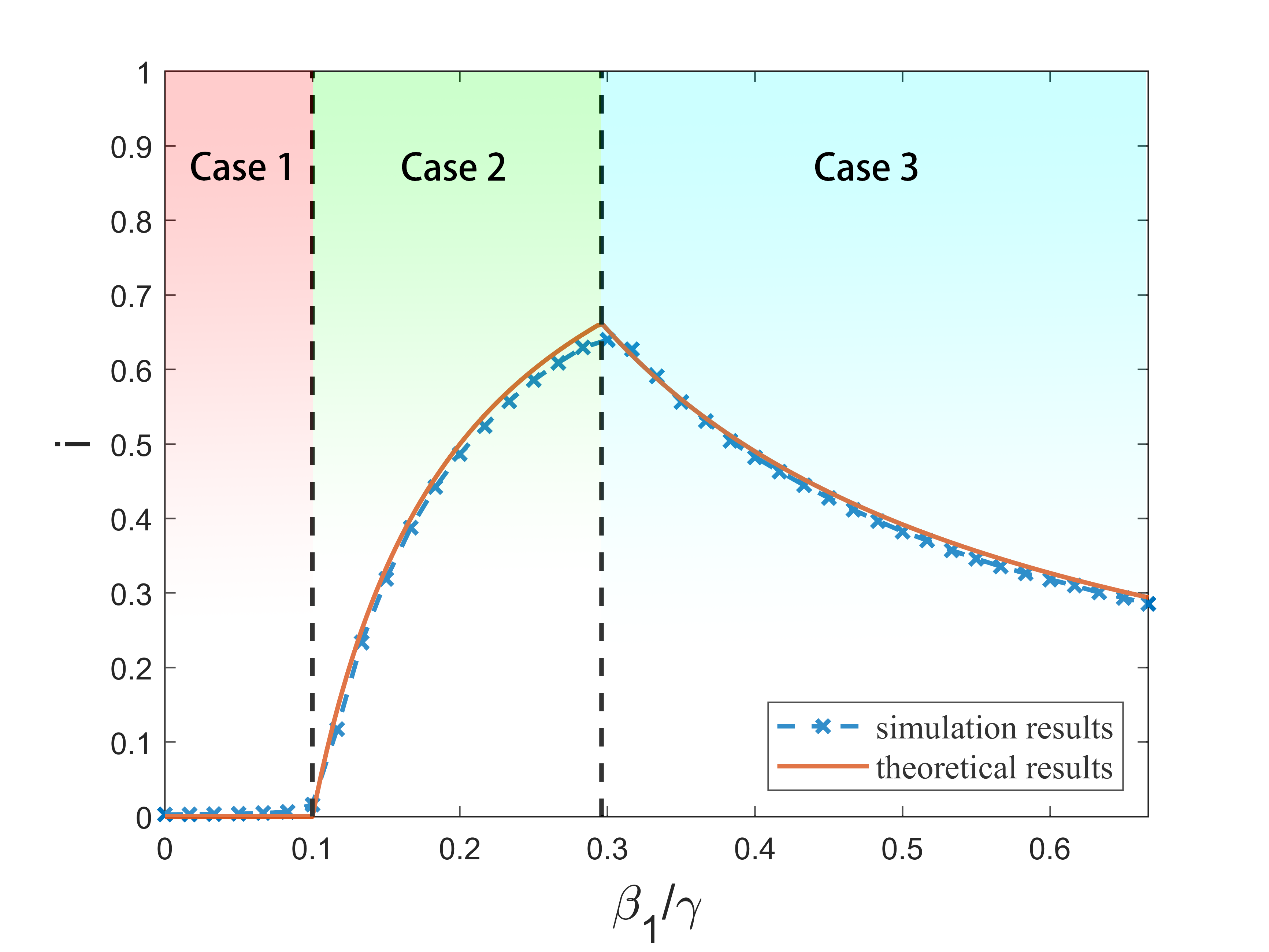}
			\label{EUTi}
		\end{minipage}%
	}%
	\subfigure[rational: $x_1^E$]{
		\begin{minipage}[t]{0.5\linewidth}
			\centering
			\includegraphics[width=0.85\linewidth]{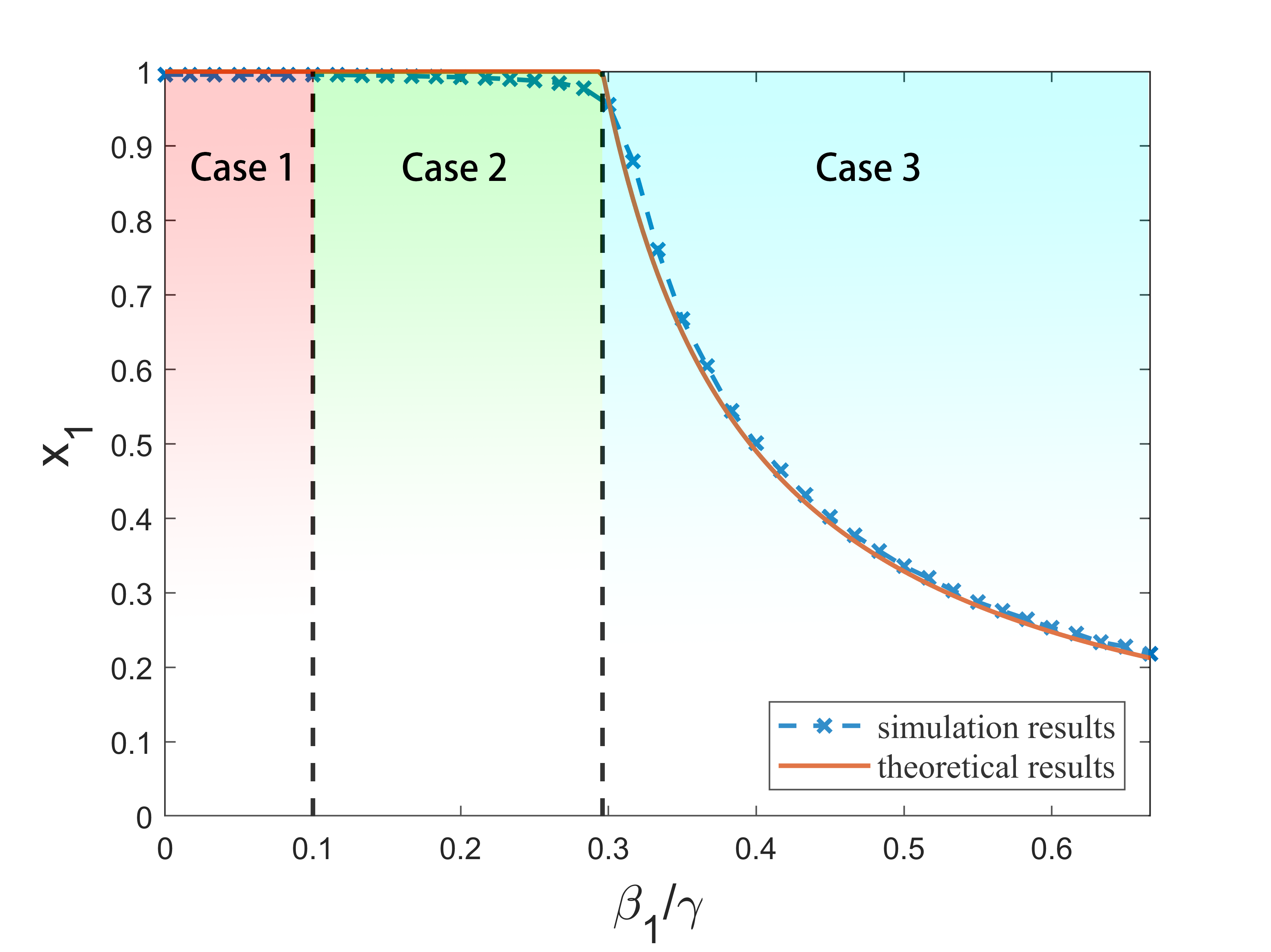}
			\label{EUTx}
		\end{minipage}%
	}%
	
	\subfigure[irrational: $i^P$]{
		\begin{minipage}[t]{0.5\linewidth}
			\centering
			\includegraphics[width=0.85\linewidth]{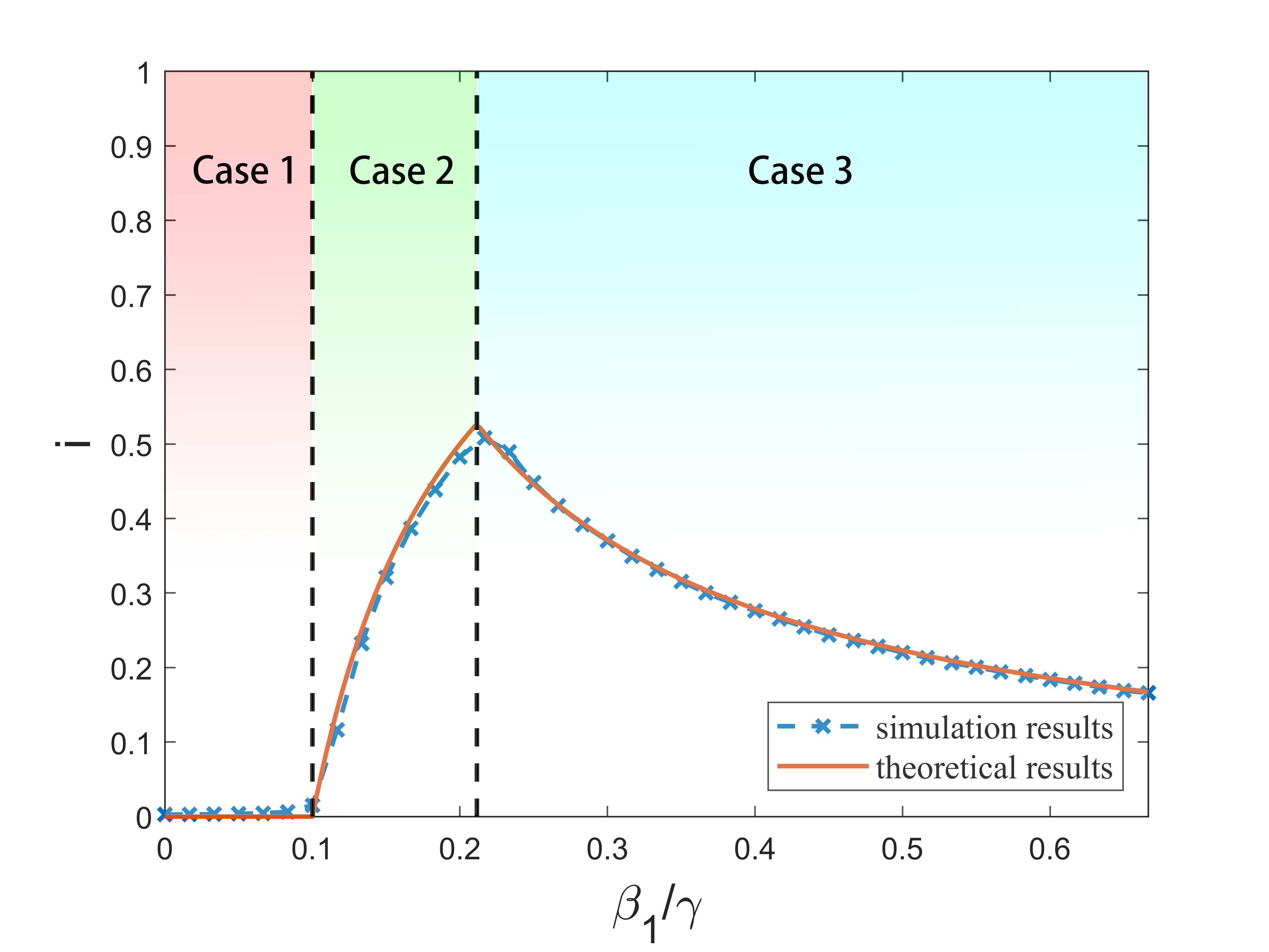}
			\label{PTi}
		\end{minipage}%
	}%
	\subfigure[irrational: $x_1^P$]{
		\begin{minipage}[t]{0.5\linewidth}
			\centering
			\includegraphics[width=0.85\linewidth]{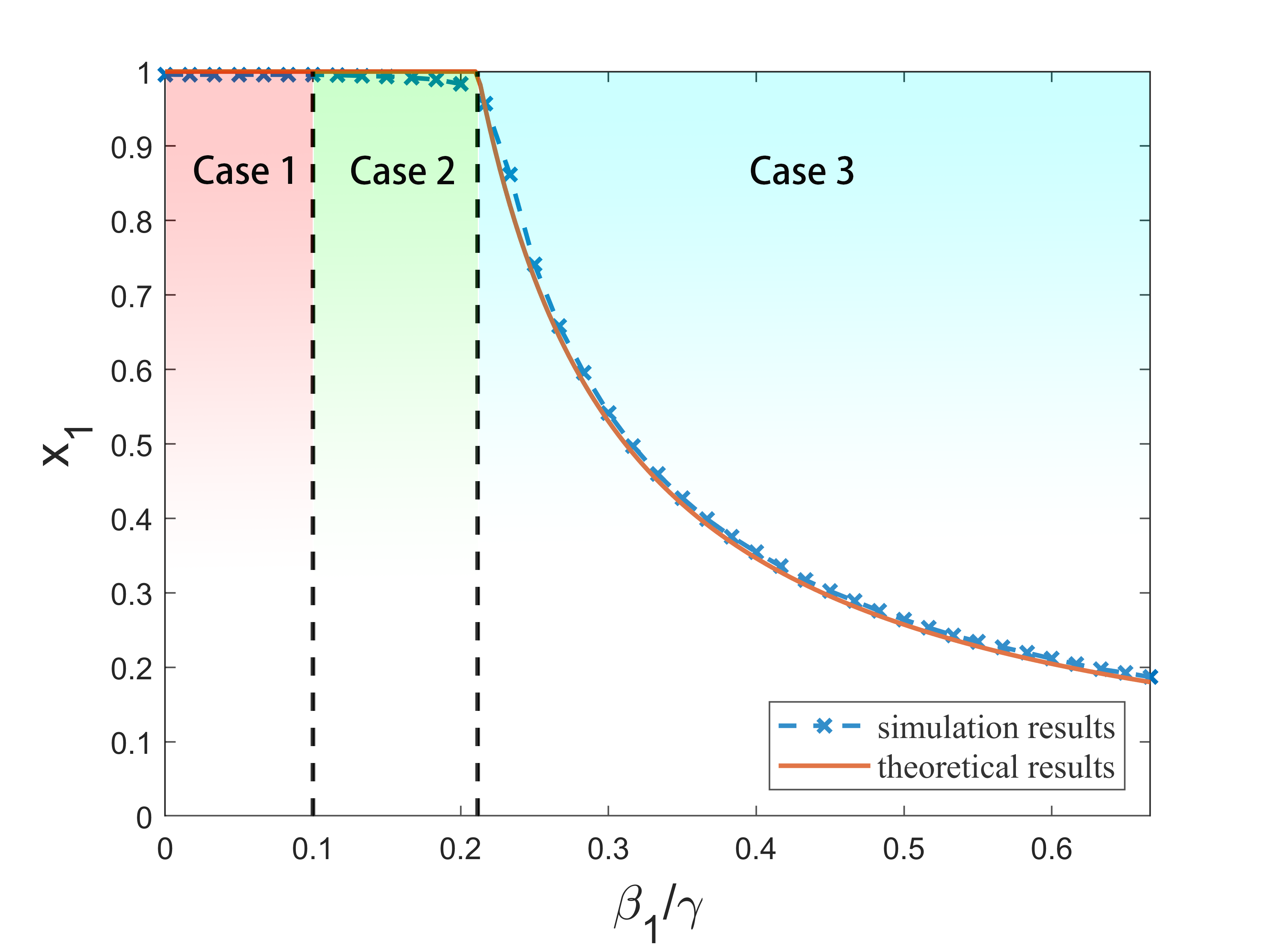}
			\label{PTx}
		\end{minipage}%
	}%
	
	\centering
	\caption{Simulation results of the steady states with rational and irrational individuals.
	}
	\label{EUT_PT}
\end{figure}

Fig. \ref{EUT_PT} shows simulation results of $(i^E,x_1^E)$, and $(i^P, x_1^P)$ with different infection rates $\beta_1$. We can see that the simulation results match the theoretical results of the steady states in Theorem \ref{th1} and \ref{th2} very well. In Fig. \ref{EUT_PT} we can clearly see that the steady states can be divided into three segments, representing the three cases. The red area is Case 1, when the infection rate is too low, all individuals choose risky behavior, and the disease does not spread. The green area is Case 2, where the proportion of infected people increases gradually as the infection rate increases. However, since the infection rate is not high enough while the payoff of risky behavior is still high, all individuals still adopt risky behavior. The blue area is Case 3 where the infection rate continues to increase, the payoff of risky behavior gradually decreases, and more individuals adopt conservative behavior. Thus, the percentage of infected individuals decreases. 

\begin{figure*}[htbp]
	
	\centering

	\subfigure{
 \scriptsize{(a)}
		\begin{minipage}[t]{0.5\linewidth}
			\centering
			\includegraphics[width=0.9\linewidth]{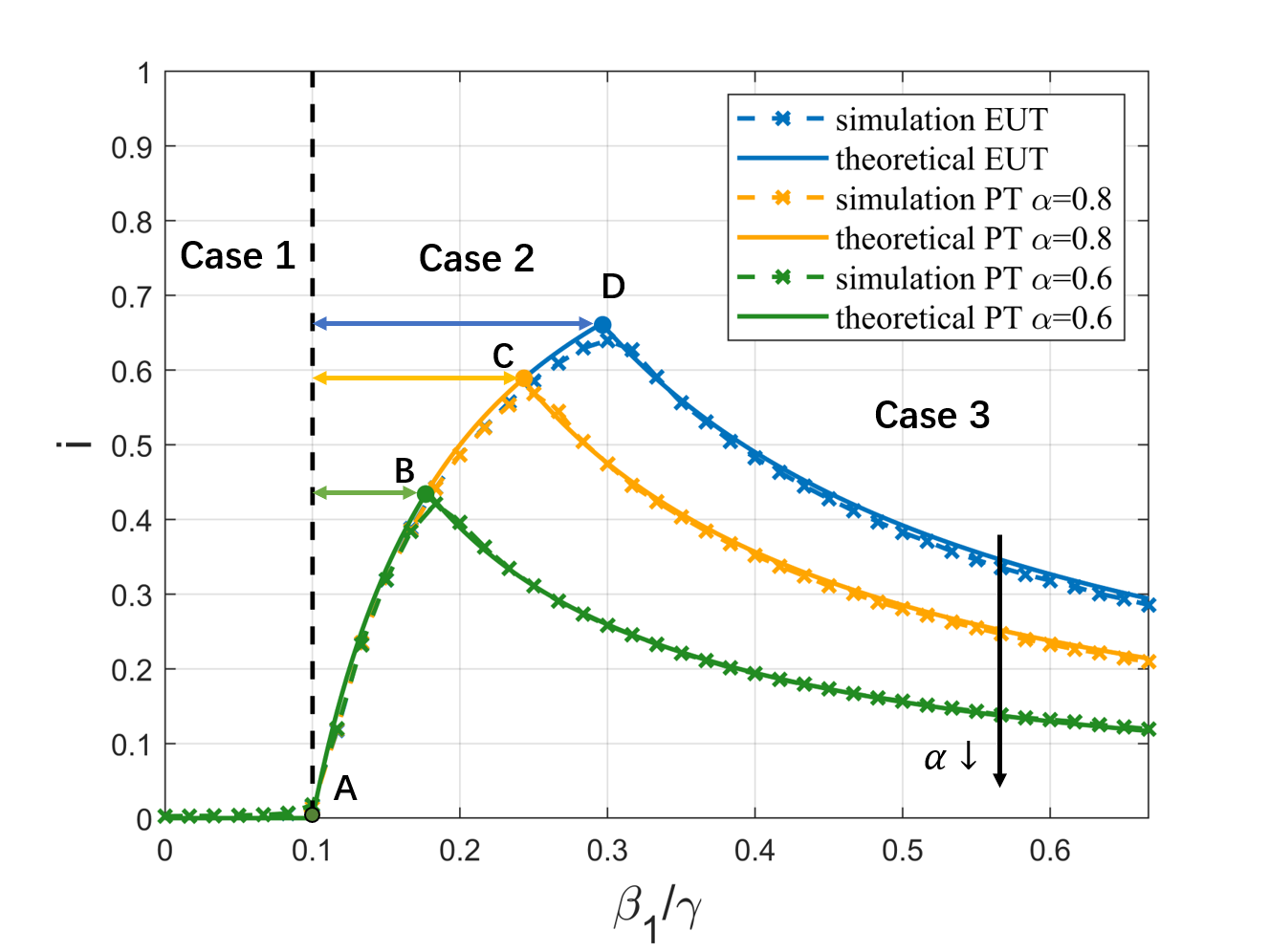}
			\label{change_beta_low1}
		\end{minipage}%
	}%
	\subfigure{
 \scriptsize{(b)}
		\begin{minipage}[t]{0.5\linewidth}
			\centering
			\includegraphics[width=0.9\linewidth]{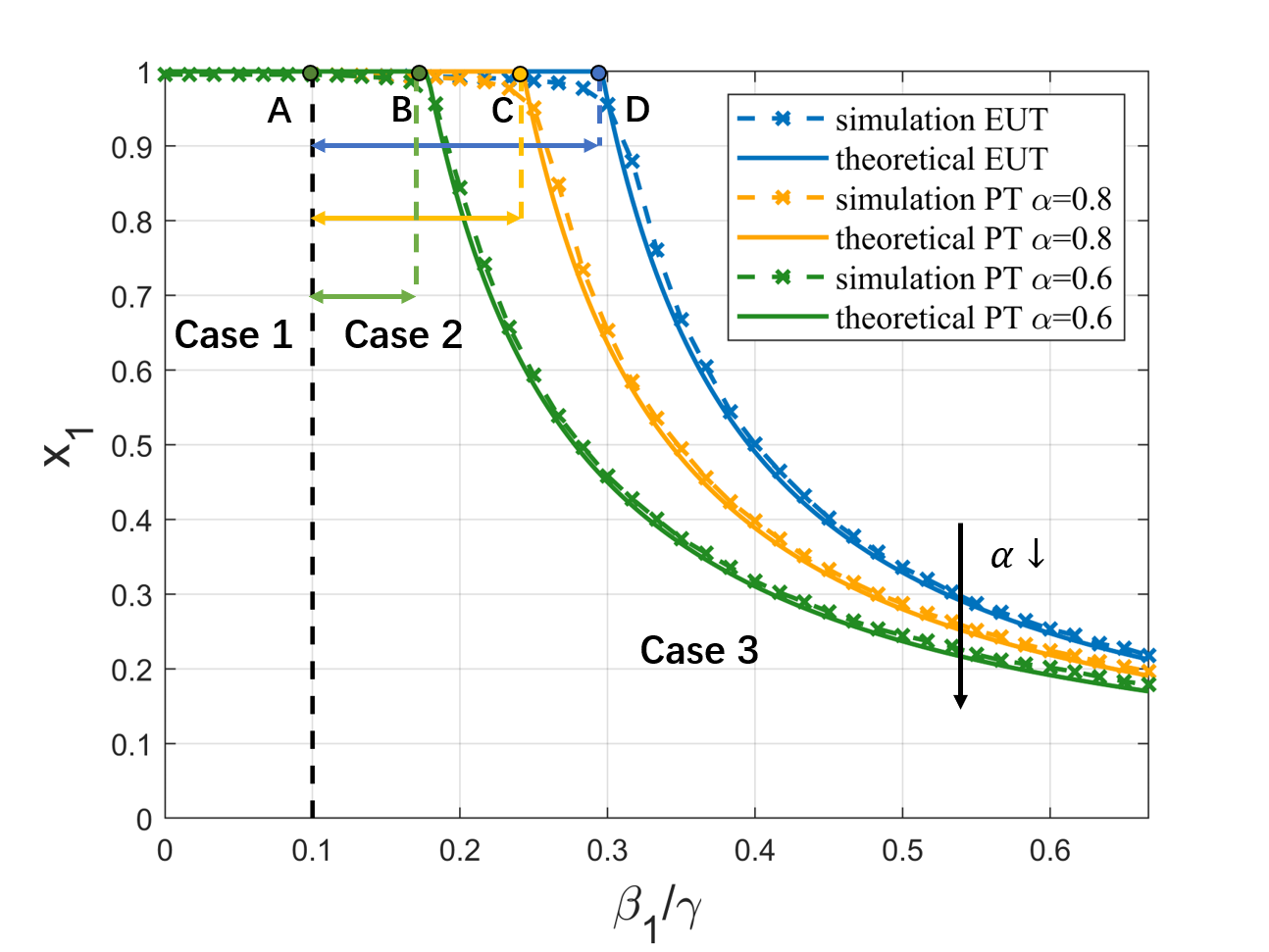}
			\label{change_beta_low2}
		\end{minipage}%
	}%
	
	\centering
	\caption{Simulation results of the steady states with different $\frac{\beta_1}{\gamma}$. (a) $i^E$ and $i^P$ (b) $x_1^E$ and $x_1^P$  
	}
	\label{change_beta_alpha}
\end{figure*}

\label{subsec:simu_irrationality}

Then we validate Theorem \ref{th4} and study the impact of irrationality on the steady states using the same simulation setup as before. 
Fig. \ref{change_beta_alpha} shows the average results of 50 simulation runs. We observe the same trend for other types of networks and other parameter settings. Note that in Fig. \ref{change_beta_alpha}, $\alpha=1$ (EUT) corresponds to the scenario with rational users, as explained in Section \ref{EGT}. In Fig. \ref{change_beta_alpha}, as $\frac{\beta_1}{\gamma}$ increases, Point A is boundary point separating Case 1 and 2, and Point B, C, and D are the boundary points separating Case 2 and 3 with $\alpha=0.6$, $\alpha = 0.8$, and $\alpha = 1$ (EUT), respectively. 
Then we analyze the results of $\alpha=0.6$ and $\alpha = 0.8$ as an example. We use $(\underline{i}^P, \underline{x}_1^P)$ and $(\bar{i}^P, \bar{x}_1^P)$ to represent the steady states for $\alpha=0.6$ and $\alpha = 0.8$. From Fig. \ref{change_beta_low1} and \ref{change_beta_low2},
\begin{itemize}
 \item before Point A, that is, when $\frac{\beta_1}{\gamma}<\frac{1}{\bar{k}}$, irrationality does not influence users' behavior and the steady states in Case 1, as both curves with $\alpha=0.6$ and $\alpha = 0.8$ have $\bar{x}_1^P = \underline{x}_1^P=1$ and $\bar{i}^P = \underline{i}^P=0$. Also, the boundary points separating Case 1 and Case 2 (Point A in Fig. \ref{change_beta_low1} and \ref{change_beta_low2}) are the same with different $\alpha$. The results validate Theorem \ref{th4}a.
 \item Between Points A and B, $\frac{\beta_1}{\gamma} > \frac{1}{\bar{k}}$, $\Phi_2 < 0$ for both $\alpha=0.6,0.8$. Therefore, from Theorem \ref{th4}b, both individuals with low irrationality ($\alpha=0.8)$ and individuals with high irrationality ($\alpha=0.6)$ take risky behavior with $\underline{x}_1^P=\bar{x}_1^P=1$ while the disease spreads to the maximum extent with $\underline{i}^P=\bar{i}^P=1-\frac{\gamma}{\bar{k}\beta_1}$, and this corresponds to Case 2 for both $(\underline{i}^P, \underline{x}_1^P)$ and $(\bar{i}^P, \bar{x}_1^P)$. 
 \item Between Points B and C, $\frac{\beta_1}{\gamma} > \frac{1}{\bar{k}}$, $\Phi_2 < 0$ for $\alpha=0.8$ but $\Phi_2 \ge 0$ for $\alpha=0.6$. From Theorem \ref{th4}b, all individuals with low irrationality still choose risky behavior with the steady state $(\bar{i}^P = 1-\frac{\gamma}{\bar{k}\beta_1}, \bar{x}_1^P = 1)$ in Case 2; while individuals with high irrationality are risk-averse with $\underline{x}_1^P<1$, and $(\underline{i}^P, \underline{x}_1^P)$ is in Case 3.
 \item After Point C, $\frac{\beta_1}{\gamma} > \frac{1}{\bar{k}}$, $\Phi_2 \ge 0$ for both $\alpha=0.8$ and $\alpha=0.6$. From Theorem \ref{th2}, both steady states for $\alpha=0.8$ and $\alpha=0.6$ are in Case 3. When comparing $(\underline{i}^P, \underline{x}_1^P)$ with $(\bar{i}^P, \bar{x}_1^P)$, as $\bar{i}^P\le\frac{1}{\bar{k}\beta_1e}$, from Theorem \ref{th4}c, irrationality makes people risk-averse, and more individuals with high irrationality choose conservative behavior with $\underline{x}_1^P<\bar{x}_1^P$, and the epidemic spreads to a smaller range with $\underline{i}^P<\bar{i}^P$.
\end{itemize}


The above shows an example where irrationality promotes conservative behavior. We find in most parameter settings, irrationality would make individuals more conservative. \ref{appendix:radical} shows an example of irrationality that makes individuals risk-seeking. It only occurs when the infection rate is high, and the loss of disease is extremely low, we analyze this situation in \ref{appendix:radicalwhy}. 
In summary, our simulation results validate our analysis in Theorem \ref{th4} when comparing the steady states with different irrationality coefficients.

\subsection{Simulations of the Behavior Inducement Algorithms}
Then we simulate the behavior inducement algorithms in Section \ref{sec:control}. We run simulations on regular networks with 500 nodes and a fixed degree of 10 for both the physical contact network and the information network. We let $c_1=0.5$, $c_2=-1$, $c_n=-10$, $\gamma=0.03$ and $\beta_1 = 0.01, 0.02$. For the loss term $l_3(\pmb{\delta})$, we use the simple 2-norm $l_3(\pmb{\delta}) = \Vert\pmb{\delta}\Vert_2^2$ as an example, and observe similar trends for other loss functions. Shown below are results averaged over 20 simulation runs.



\begin{figure}[htbp]
	
	\centering
	\subfigure[$\frac{\beta_1}{\gamma}=\frac{2}{3}$]{
		\begin{minipage}[t]{0.5\linewidth}
			\centering
			\includegraphics[width=0.9\linewidth]{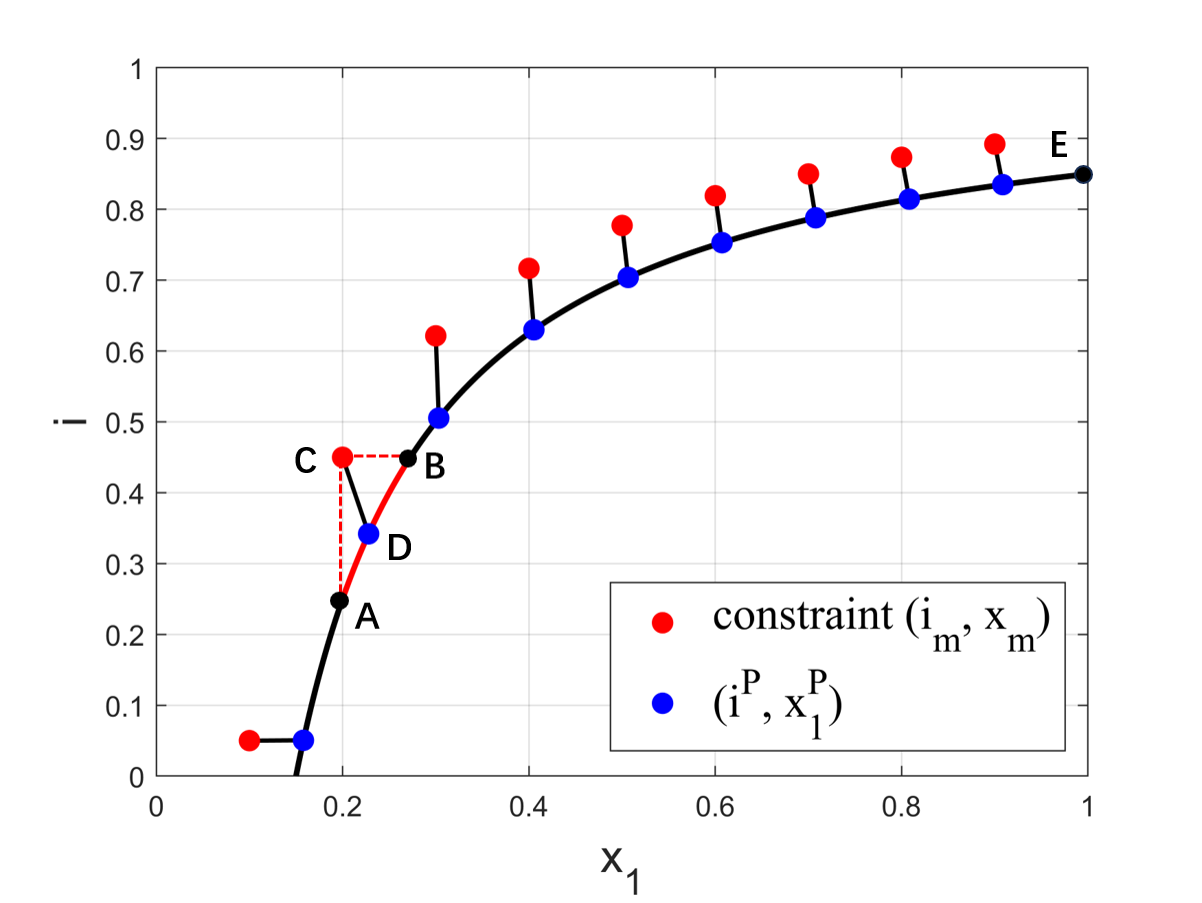}
			\label{optimize_out1}
		\end{minipage}%
	}%
	\subfigure[$\frac{\beta_1}{\gamma}=\frac{1}{2}$]{
		\begin{minipage}[t]{0.5\linewidth}
			\centering
			\includegraphics[width=0.9\linewidth]{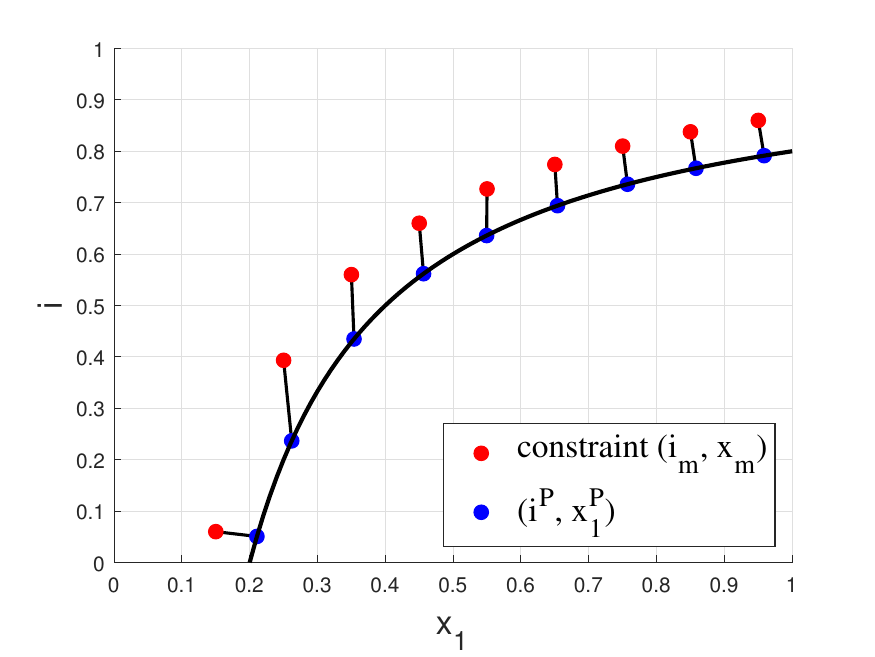}
			\label{optimize_out2}
		\end{minipage}%
	}%
	
	\centering
	\caption{Simulation results of behavior inducement algorithm with feasible constraints $(i_m, x_m)$. 
	}
	\label{optimize_out}
\end{figure}

\begin{figure}[htbp]
	
	\centering
	\subfigure{
 \scriptsize{(a)}
		\begin{minipage}[t]{0.5\linewidth}
			\centering
			\includegraphics[width=0.9\linewidth]{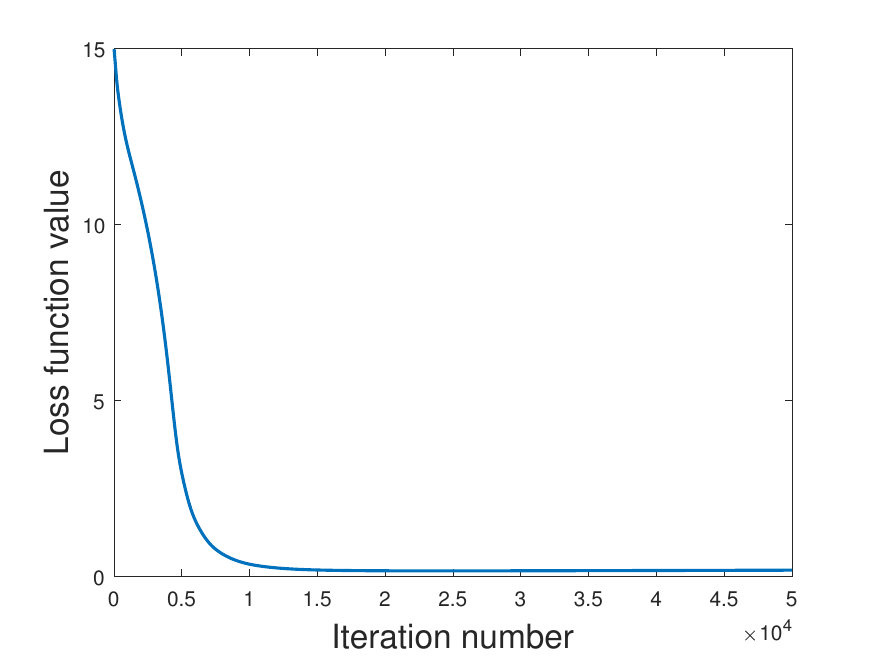}
			\label{optimize_loss1}
		\end{minipage}%
	}%
	\subfigure{
 \scriptsize{(b)}
		\begin{minipage}[t]{0.5\linewidth}
			\centering
			\includegraphics[width=0.9\linewidth]{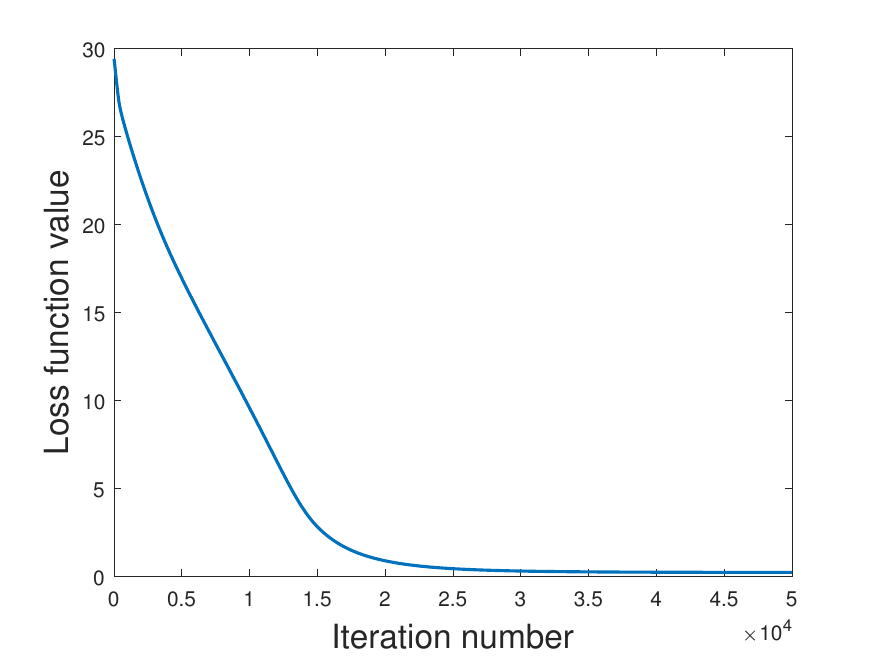}
			\label{optimize_loss2}
		\end{minipage}%
	}%
	
	\centering
	\caption{Simulation results of the behavior inducement algorithms. (a) The loss function in (\ref{eq:bound}) with feasible constraints, and (b) the loss function in (\ref{eq:unbound}) with infeasible constraints.}
	\label{optimize_loss}
\end{figure}

The results of the proposed algorithm with feasible constraints are shown in Fig. \ref{optimize_out}. The black solid line is the boundary between the feasible and infeasible regions, where $(i_m, x_m)$ above this line are feasible constraints where it is possible to find $\pmb{\delta}$ to satisfy $0\le i^P \leq i_m$ and $1\ge x_1^P \geq x_m$, and $(i_m, x_m)$ below this line are infeasible. Moreover, since $\beta_1$, $\gamma$ and $\bar{k}$ are fixed, the black solid line includes all possible $(i^P, x_1^P)$, where point E represents Case 2 and the other points in the black solid line represent Case 3. The red dots represent the limits $(i_m, x_m)$ we set in our simulations, and it can be seen that they are all feasible. The blue dots show the steady states after using the proposed behavior guidance algorithm in Section \ref{sec:boundedconstraints}, and they all satisfy $0\le i^P \leq i_m$ and $1\ge x_1^P \geq x_m$. Using the constraint $(i_m, x_m)$ labeled as \textbf{Point C} in Fig. \ref{optimize_out1}, the steady state $(i^P, x_1^P)$ should fall within the range illustrated from point A to point B to satisfy the constraints $0\le i^P \leq i_m$ and $1\ge x_1^P \geq x_m$, and the proposed behavior guidance algorithm selects \textbf{Point D} in this range that minimizes the loss function in \eqref{eq:bound}. The values of the loss function in \eqref{eq:bound} at each iteration are shown in Fig. \ref{optimize_loss1}. As shown there, the loss function decreases after each iteration and converges after 10,000 iterations.

\begin{figure}[htbp]
	
	\centering
	\subfigure[$\frac{\beta_1}{\gamma}=\frac{2}{3}$]{
		\begin{minipage}[t]{0.5\linewidth}
			\centering
		
  \includegraphics[width=0.9\linewidth]{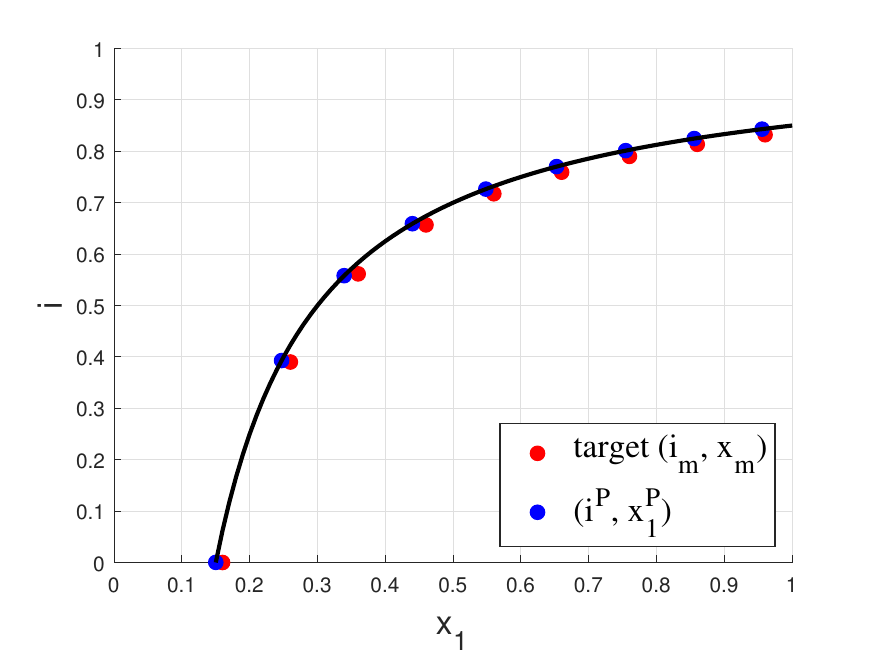}
			\label{optimize_in1}
		\end{minipage}%
	}%
	\subfigure[$\frac{\beta_1}{\gamma}=\frac{1}{2}$]{
		\begin{minipage}[t]{0.5\linewidth}
			\centering
			\includegraphics[width=0.9\linewidth]{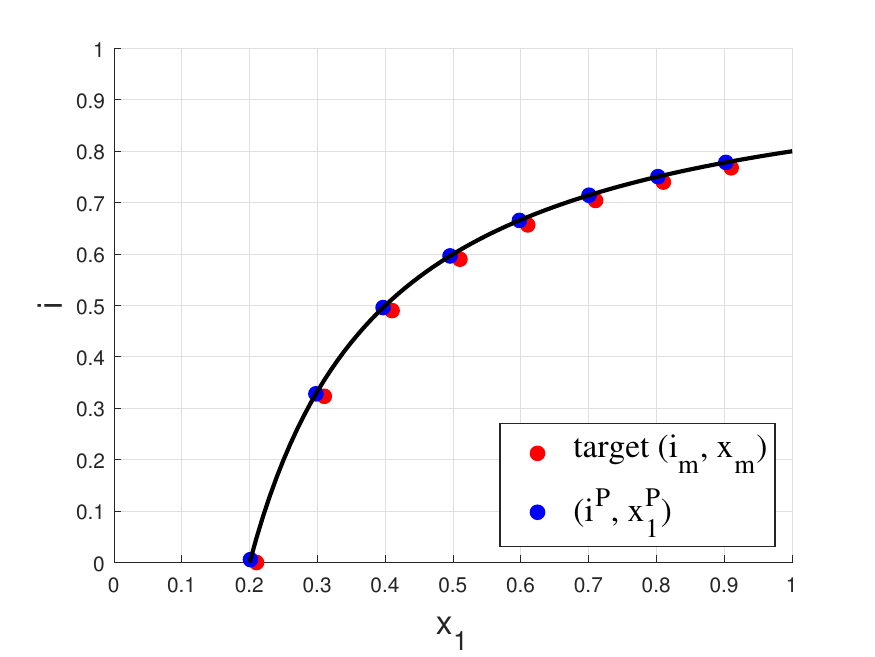}
			\label{optimize_in2}
		\end{minipage}%
	}%
	
	\centering
	\caption{Simulation results of the behavior inducement algorithm with infeasible constraints $(i_m, x_m)$.
	}
	\label{optimize_in}
\end{figure}

Simulation results of the algorithm with infeasible constraints are shown in Fig.\ref{optimize_in}. Here, the given two constraints ($i_m$, $x_m$) are below the boundary of the feasible region and impossible to satisfy at the same time. It can be seen that although the steady state with behavior guidance cannot satisfy the two constraints simultaneously, it is on the boundary of the feasible region and very close to $(i_m, x_m)$. The value of the loss function in \eqref{eq:unbound} after each iteration is shown in Fig. \ref{optimize_loss2}, where we can see that it decreases and converges as the number of iterations increases. 

\section{Real User Tests}
\label{sec:real}
While the simulation experiments have validated our theoretical analysis, it is necessary to further validate our conclusions using real user tests. However, obtaining real user data on behavioral choices, network structure, and spread data simultaneously poses a significant challenge. Therefore, we qualitatively validate our conclusions through sociological experiments. In our test, 141 subjects are interviewed, including 61 males and 80 females. Their occupations include students, production staff, salespersons, human resources, teachers, etc. Their ages range from 18 to 60. 





In our test, we first collect data to estimate the irrationality coefficient $\alpha$ of the subjects. Note that in Section \ref{sec:Model} and \ref{sec:analysis}, to simplify the analysis, we assume that all individuals have the same $\alpha$, which can be considered as the average of $\alpha$ over the entire network. In reality, the irrationality coefficient $\alpha$ varies from person to person. Therefore, in our test, we estimate $\alpha$ for each individual separately. Next, we gather data on the subjects' behavioral choices in various scenarios during a pandemic. This data collection process allows us to capture individuals' risk preferences. Finally, we analyze the relationship between the irrationality coefficient $\alpha$ and the risk preference. By examining this relationship, we gain insights into how individuals' irrationality impacts their decisions during a pandemic and validate our analysis in Section \ref{sec:analysis}.

\subsection{Estimating the Irrationality Coefficient $\alpha$}
\textbf{Data collection: }Following the works in \cite{gonzalez1999shape,prelec2000compound,tversky1992advances}, we estimate the subjects' irrationality coefficient $\alpha$ in a way similar to gambling games. In our experiments, subjects are presented with a scenario where they have a probability $p_i$ of incurring a large financial loss. However, they are also given the option to purchase insurance at different prices to mitigate the potential loss. The subjects are then asked to make a decision regarding whether they would choose to buy the insurance. Below is an example we use in our experiment.

\emph{Question A: You have a probability $10\%$ of losing \textyen100, but if you choose to buy insurance, you can guarantee that you will not bear this loss. Then what is the insurance price you can accept?}

	\emph{a}. \emph{When the price is lower than \textyen10, I would buy it.}
 
	\emph{b}. \emph{When the price is lower than \textyen20 but higher than \textyen10, I would buy it.}
	
	\emph{c}. \emph{When the price is lower than \textyen30 but higher than \textyen20, I would buy it.}
	
	\emph{d}. \emph{When the price is lower than \textyen40 but higher than \textyen30, I would buy it.}
	
	\emph{e}. \emph{When the price is lower than \textyen50 but higher than \textyen40, I would buy it.}
	
	\emph{f}. \emph{Even if the price is higher than \textyen50, I would buy it.}

After the subjects have made their initial choices, a refined set of choices is presented to them to obtain fine-grained results. For instance, if a subject chose option $c$ in the above question, the prices shown in the subsequent question would be narrowed down to a range between \textyen30 and \textyen40, such as \{\textyen30-\textyen32, \textyen32-\textyen34, \textyen34-\textyen36, \textyen36-\textyen38, \textyen38-\textyen40\}. This process continues until the range is narrowed down to \textyen$1$.
For each subject, we change the probability $p_i$ in the questionnaire, repeat the above process, and get multiple pairs of $(p_i, r_i)$, where $r_i$ is the final acceptable insurance price of the subject.

\textbf{Estimation of $\alpha$:} If the subject chooses an acceptable insurance price of $r_i$ to avoid the loss of \textyen100 with probability $p_i$, then for this subject, a loss of \textyen100 with probability $p_i$ is equivalent to a loss of $r_i$ with probability 1. Then from \eqref{U_PT}, we have:

\begin{equation}
\begin{split}
\pmb{\omega}(p_i,\alpha)u^P(-100) +\pmb{\omega}(1-p_i,\alpha)u^P(0)= \pmb{\omega}(1,\alpha)u^P(-r_i).
\end{split}
\label{51}
\end{equation}

Similar to \cite{prelec2000compound}, we use (\ref{eqn:probweightfunc}) and (\ref{utilityfunc}) as the probability weighting function and the value function, respectively, in our work. Note that from the definitions, we have $u^P(0)=0$ and $\pmb{\omega}(0,\alpha)=0$, then we have
\begin{equation}
\begin{split}
-e^{(-(-lnp_i)^\alpha)}\lambda (100)^\sigma = -\lambda (r_i)^\sigma,
\end{split}
\label{53}
\end{equation}
which is equivalent to
\begin{equation}
ln(-ln(\frac{r_i}{100})) = \alpha ln(-ln(p_i)) - ln(\sigma).
\label{54}
\end{equation}

In \eqref{54}, $ln(-ln(\frac{r_i}{100}))$ is a linear function of $ln(-ln(p_i))$.  We set $\sigma=0.65$ following the work in \cite{prelec2000compound}. Given the collected pairs $\{ (p_i, r_i) \}$ from one subject, we use linear regression to find $\alpha$ in \eqref{54} for this subject. 

\subsection{Measuring Individuals' Risk Preference}
\textbf{Data collection:} Then we proceed to collect data on the behavioral choices of different subjects when confronted with risky scenarios during a pandemic. In each question of this section, subjects are presented with a specific epidemic situation. Within each scenario, subjects must choose between going out and staying at home. They are informed that if they choose to go out, there is a certain probability of becoming infected. On the other hand, if they decide to stay at home, they are guaranteed not to be infected, but they will experience some form of loss. Below is an example in our experiment.

\emph{Question B: There is an epidemic spreading right now. If you come into contact with an infected person, you have a 5\% chance of being infected. If you choose to go out, you will be in close contact with 20 people every day. If you decide to stay at home, you are guaranteed to avoid infection but will experience some losses. On the other hand, if you go out, there is a chance of becoming infected. The loss of being infected is 20 times that of the loss due to home isolation. Your city has a population of 1 million, then:}

	\emph{a}. \emph{When there is no confirmed case in the city, I will go out.}
	
	\emph{b}. \emph{When the number of confirmed cases in the city is less than 10, I will go out.}
	
	\emph{c}. \emph{When the number of confirmed cases in the city is less than 100, I will go out.}
	
	\emph{d}. \emph{When the number of confirmed cases in the city is less than 1,000, I will go out.}
	
	\emph{e}. \emph{When the number of confirmed cases in the city is less than 10,000, I will go out.}
	
	\emph{f}. \emph{When the number of confirmed cases in the city is less than 100,000, I will go out.}
	
	\emph{g}. \emph{When the number of confirmed cases in the city is higher than 100,000, I will still go out.}


\textbf{Calculating Individuals' Risk Preference $I_x$:} We use the risk preference $I_x$ to reflect the behavioral tendencies of the subjects. 
The risk preference $I_x$ for each individual corresponds to the proportion of times choosing to engage in risky behavior in various scenarios in the questionnaire. This indicator ranges between 0 and 1, where values closer to 1 indicate a more risky behavioral tendency. It is important to note that in our model, $x_1^P$ denotes the proportion of individuals choosing the risky behavior. A higher $I_x$ value within a group implies a greater inclination of individuals towards risky behavior, resulting in a higher $x_1^P$ value. Then we analyze the relationship between the irrationality and behavioral choice.

\begin{figure}[t]
	\centering
		\begin{minipage}[t]{0.5\linewidth}
			\centering
			\includegraphics[width=1\linewidth]{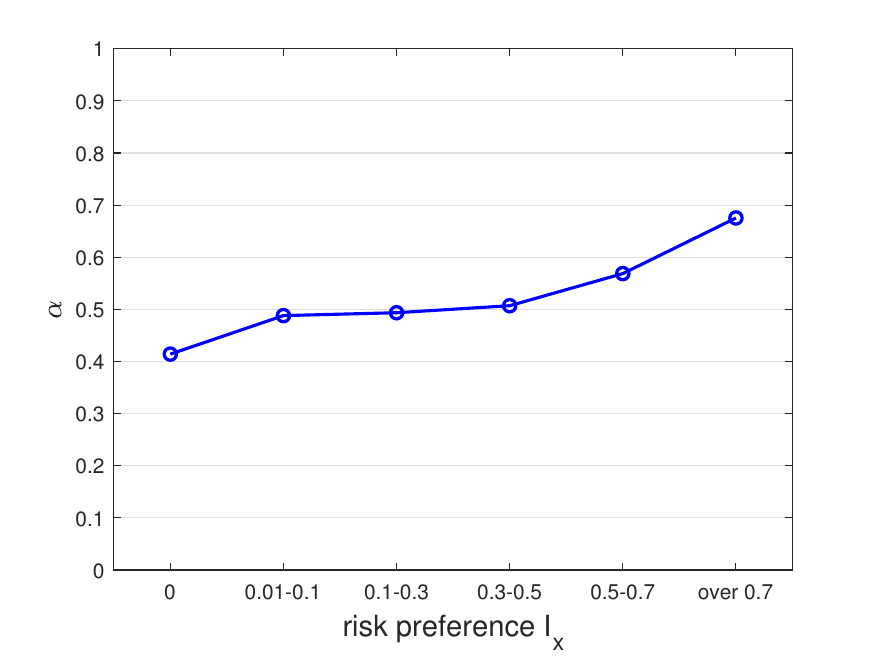}
			\label{fig:social1}
		\end{minipage}%
	\centering
	\caption{The relationship between risk preference $I_x$ and $\alpha$.}
	\label{fig:social}
\end{figure}

\subsection{Analysis of the Relationship Between Irrationality and Behavioral Choice}
We classify subjects into six groups based on their risk preference, with the aim of grouping together individuals who share similar values of $I_x$ within each respective group. We calculate the mean value of $\alpha$ and the average of $I_x$ for each group and analyze the relationship between $\alpha$ and $I_x$. The results are illustrated in Fig. \ref{fig:social}. We can see that groups with a smaller $\alpha$ (i.e., a higher degree of irrationality) also have a smaller average risk preference $I_x$, which is consistent with our conclusion. In our theoretical analysis and simulations, we find that irrationality makes more conservative in most cases, and irrationality makes individuals more risk-seeking only occurs when the infection rate is high, and the loss of disease is extremely low. Since the parameter settings of our real user tests do not meet this condition. Therefore, irrationality leads individuals to be more conservative and has a smaller risk preference.

To validate the validity of our theory from a statistical point of view, we get the Pearson correlation coefficient and the Spearman correlation coefficient of the $I_x$ and $\alpha$ (For the Pearson correlation coefficient, we take the mean of $I_x$ for each group as a variable to calculate). The results are shown in Table \ref{pearson}. Both the Pearson correlation coefficient and the Spearman correlation coefficient reveal a significant positive correlation between $I_x$ and $\alpha$. This indicates that irrationality leads individuals to exhibit more conservative behavior.

\begin{table}[htbp] 
\captionsetup[table]{position=top}
	\renewcommand\arraystretch{1.2}
	\centering  
	\caption{Pearson correlation coefficient}   
	\begin{tabular}{l|cccc} 
		\toprule 
		Parameters &  Pearson correlation & Significance & Spearman correlation & Significance\\
		\midrule 
		$I_r$ and $\alpha$   & 0.939 &  0.005 & 1.000 &  0.000         \\

		\bottomrule 
	\end{tabular}
	\label{pearson}
\end{table}


\color{black}


\section{Conclusion}
\label{sec:conclusion}
In this paper, we propose an epidemic-behavior coevolution framework to analyze the behavior of individuals and the coevolution of user behavior and the disease spread during an epidemic. Our model takes into account the irrationality of individuals' decision-making processes, and our theoretical analysis shows that individual irrationality polarizes individual behavior choices. That is, irrationality makes users risk-averse when the probability of being infected is small, while they tend to be risk-seeking when the probability of being infected is large. We then propose a behavior inducement algorithm to control the disease spread and reduce losses by guiding individual behavior. Simulation results show the correctness of our theoretical analysis and verify the validity of our guidance control method. We also qualitatively prove the correctness of our conclusions using real-user tests.

\bibliography{mylib}
 \clearpage
\appendix

\section{The proof of Theorem \ref{th1}}
\label{appendix:th1}
In order to analyze the steady state, we first introduce the Lemma \ref{lemma1}.
\begin{lemma}
	For the 2-behavior model, the necessary and sufficient conditions that $(i^*,x_1^*)$ is a steady state is:
	\begin{equation}
	\begin{split}
	&\left.\dfrac{di}{dt}\right|_{i=i^*} = 0,\left.\dfrac{dx_1}{dt}\right|_{x_1=x_1^*} =0,\\
	&\left.(\dfrac{\partial i'}{\partial i}+\dfrac{\partial x_1'}{\partial x_1})\right|_{i=i^*,x_1=x_1^*}<0,\\
	&\left.(\dfrac{\partial i'}{\partial i}\dfrac{\partial x_1'}{\partial x_1}-\dfrac{\partial i'}{\partial x_1}\dfrac{\partial x_1'}{\partial i})\right|_{i=i^*,x_1=x_1^*}>0.
	\end{split}
	\end{equation}
	\label{lemma1}
\end{lemma}

\emph{Proof}: By the definition of steady state, we first have $\dfrac{di}{dt} = 0$ and $\dfrac{dx_1}{dt} =0$.  For simplicity, we set 
\begin{equation}
    \begin{aligned}
        &P(i^*,x_1^*) = \left.(\dfrac{\partial i'}{\partial i}+\dfrac{\partial x_1'}{\partial x_1})\right|_{i=i^*,x_1=x_1^*},\\
        &Q(i^*,x_1^*) = \left.(\dfrac{\partial i'}{\partial i}\dfrac{\partial x_1'}{\partial x_1}-\dfrac{\partial i'}{\partial x_1}\dfrac{\partial x_1'}{\partial i})\right|_{i=i^*,x_1=x_1^*}.
    \end{aligned}
\end{equation}

Then we have:
\begin{equation}
\lambda_1=\frac{P+\sqrt{P^2-4Q}}{2},\ \lambda_2=\frac{P-\sqrt{P^2-4Q}}{2}
\end{equation}

Since $Re(\lambda_1)<0$ and $Re(\lambda_2)<0$, we have $P<0$ and $Q>0$, which means $\frac{\partial i'}{\partial i}+\frac{\partial x_1'}{\partial x_1}<0$ and $\frac{\partial i'}{\partial i}\frac{\partial x_1'}{\partial x_1}-\frac{\partial i'}{\partial x_1}\frac{\partial x_1'}{\partial i}>0$.

Then we proof the Theorem \ref{th1} based on the Lemma 1:

By setting $\frac{di}{dt} = 0,\frac{dx_1}{dt} =0$, we can get four points: $(0,0)$, $(0,1)$, $(1-\frac{\gamma}{\bar{k}\beta_1},1)$ and $\left(\frac{k_0(u^{E}(c_2)-u^{E}(c_1))}{(k_0u^{E}(c_n)-1)\bar{k}\beta_1},\frac{(k_0u^{E}(c_n)-1)\gamma}{\bar{k}\beta_1(k_0u^{E}(c_n)-1)-k_0(u^{E}(c_2)-u^{E}(c_1))}\right)$. Then we discuss the stability of these points. Based on Lemma \ref{lemma1}, the steady state should satisfy $P(i,x_1) = \frac{\partial i'}{\partial i}+\frac{\partial x_1'}{\partial x_1}<0$ and $Q(i,x_1) = \frac{\partial i'}{\partial i}\frac{\partial x_1'}{\partial x_1}-\frac{\partial i'}{\partial x_1}\frac{\partial x_1'}{\partial i}>0$. In our model, the individual can adopt two behaviors, $a_1$ for risky behavior (like going out) and $a_2$ for conservative behavior (like self-isolation). In general, the utility of going out should be larger than self-isolation, so we have $c_1-c_2>0$ (that means $u^{E}(c_1)-u^{E}(c_2)>0$ and $u^{P}(c_1)-u^{P}(c_2)>0$).
And $c_n$ is the loss of being infected, so $c_n<0$ (that means $u^{E}(c_n)<0$ and $u^{P}(c_n)<0$).
\begin{itemize}
	
	\item For the point $(0,0)$, we have:
	\begin{equation}
	\begin{split}
	&P=k_0(u^{E}(c_1)-u^{E}(c_2))-\gamma,\\ &Q=-\gamma k_0(u^{E}(c_1)-u^{E}(c_2)).
	\end{split}
	\label{20}
	\end{equation}
	
	Since $u^{E}(c_1)-u^{E}(c_2)>0$, $\gamma>0$ and $k_0>0$, we have $Q<0$.
	So $(0,0)$ is not a steady state.

	\item For the point $(0,1)$, we have:
	\begin{equation}
	\begin{split}
	&P=-(\gamma-\bar{k}\beta_1)-k_0(u^{E}(c_1)-u^{E}(c_2)),\\ &Q=k_0(\gamma-\bar{k}\beta_1)(u^{E}(c_1)-u^{E}(c_2)).
	\label{21}
	\end{split}
	\end{equation}
	
	Since $u^{E}(c_1)-u^{E}(c_2)>0$ and $k_0>0$, if $Q>0$, we have $\gamma-\bar{k}\beta_1>0$ (equivalent to $\bar{k}<\frac{\gamma}{\beta_1}$).  
	When $\gamma-\bar{k}\beta_1>0$, we also have $P<0$.
	Therefore, when $\bar{k}<\frac{\gamma}{\beta_1}$, $(0,1)$ is a steady state.

\item For the point $\left(1-\frac{\gamma}{\bar{k}\beta_1},1\right)$, we have:
	\begin{equation}
	\begin{aligned}
	&P=&&-k_0(u^{E}(c_1)-u^{E}(c_2))+(\gamma-\bar{k}\beta_1)k_0u^{E}(c_n),\\
	&Q=&&(\gamma-\bar{k}\beta_1)[-k_0(u^{E}(c_1)-u^{E}(c_2))+(\gamma-\bar{k}\beta_1)(k_0u^{E}(c_n)-1)].
	\end{aligned}
	\end{equation}
	
	Since $0\le i\le 1$, then $0\le 1-\frac{\gamma}{\bar{k}\beta_1}\le 1$, we have $\bar{k}>\frac{\gamma}{\beta_1}$ (if $\bar{k}=\frac{\gamma}{\beta_1}$, then $Q=0$). When $\bar{k}>\frac{\gamma}{\beta_1}$, we have $\gamma-\bar{k}\beta_1<0$. Then $Q>0$, $P<0$ is equivalent to:
	\begin{equation}
	\begin{aligned}
	&g_1=-k_0(u^{E}(c_1)-u^{E}(c_2))+(\gamma-\bar{k}\beta_1)k_0u^{E}(c_n),\\
	&g_2=-k_0(u^{E}(c_1)-u^{E}(c_2))+(\gamma-\bar{k}\beta_1)(k_0u^{E}(c_n)-1),\\
	&g_1<0,\ g_2<0.
	\end{aligned}
	\end{equation}
	
	It is obvious that $g_1=g_2-(\gamma-\bar{k}\beta_1)$. If $g_2<0$, we have $g_1<0$. Note that $g_2=\Phi_1$
	
	Therefore, when $\bar{k}>\frac{\gamma}{\beta_1}$ and $\Phi_1<0$, the point $\left(1-\frac{\gamma}{\bar{k}\beta_1},1\right)$ is a steady state.
	
	\item For the point $(i^{(1)},x_1^{(1)})$ ,we have:
 \begin{equation}
     i^{(1)}=\frac{k_0(u^{E}(c_2)-u^{E}(c_1))}{(k_0u^{E}(c_n)-1)\bar{k}\beta_1}, \  x_1^{(1)}=\frac{\gamma}{(1-i^{(1)})\bar{k}\beta_1}.
     \label{eq:ixEUT}
 \end{equation}

First, as $i^{(1)}$ and $\frac{\gamma}{(1-i^{(1)})\bar{k}\beta_1}$ are the proportion of infected individuals and the proportion of individuals who choose risky behavior, by definition, we have $0\le i^{(1)} \le 1$ and $0\le \frac{\gamma}{(1-i^{(1)})\bar{k}\beta_1} \le 1$, which is equivalent to:
	\begin{equation}
	0\le i^{(1)} \le 1-\frac{\gamma}{\bar{k}\beta_1}.
	\label{eq:condi}
	\end{equation}
	
	Since $0\le i^{(1)} \le 1-\frac{\gamma}{\bar{k}\beta_1}$, it should satisfies $0<1-\frac{\gamma}{\bar{k}\beta_1}$, which means $\bar{k}>\frac{\gamma}{\beta_1}$. (If $0=1-\frac{\gamma}{\bar{k}\beta_1}$, then the state $\left(i^{(1)},\frac{\gamma}{(1-i^{(1)})\bar{k}\beta_1}\right)$ becomes $(0,1)$, which we have discussed). 
 
	 Since $u^{E}(c_1)-u^{E}(c_2)>0$, $k_0u^{E}(c_n)-1<0$, then we have $i^{(1)}> 0$. Then if $i^{(1)} \le 1-\frac{\gamma}{\bar{k}\beta_1}$, based on \eqref{eq:ixEUT}, it should satisfy:
	\begin{equation}
	\begin{split}
	\Phi_1=-k_0(u^{E}(c_1)-u^{E}(c_2))+(\gamma-\bar{k}\beta_1)(k_0u^{E}(c_n)-1) \ge 0.
	\end{split}
	\end{equation}

	Note that $x_1^{(1)}=\frac{\gamma}{(1-i^{(1)})\bar{k}\beta_1}$, we have $P=-\bar{k}\beta_1 i^{(1)} x_1^{(1)}$.
	Since $0\le i^{(1)} \le 1$ and $0\le x_1^{(1)} \le 1$, we have $P<0$ and $Q>0$.
	
	In summary, $(i^{(1)},x_1^{(1)})$ is a steady state when $\bar{k}>\frac{\gamma}{\beta_1}$ and $\Phi_1\ge 0$.

\end{itemize}

\section{Proof for the maximum spread extent}
\label{adx:A}
For the steady state $(i^E,x_1^E)$, it satisfies $\frac{di}{dt}=0$. If $i^E\neq 0$, then based on \eqref{eq:EUT}, we have:
\begin{equation}
i^E=1-\frac{\gamma}{\bar{k}\beta_1x_1^E},
\end{equation}
and $i^E$ is an increasing function of $x_1^E$, where $x_1^E \in [0,1]$. Therefore, $i^E$ reaches its maximum value when $x_1^E=1$. Substituting this value into the equation, we obtain $\hat{i}=1-\frac{\gamma}{\bar{k}\beta_1}$.
\section{The proof of Theorem \ref{th2}}
\label{appendix:th2}
By setting $\frac{di}{dt} = 0,\frac{dx_1}{dt} =0$, we can also get four points: $(0,0)$, $(0,1)$, $\left(1-\frac{\gamma}{\bar{k}\beta_1},1\right)$ and $\left(i^{(2)},\frac{\gamma}{(1-i^{(2)})\bar{k}\beta_1}\right)$, then we discuss whether they are steady states.

\begin{itemize}
	\item For the point $(0,0)$, we have $P=k_0(u^P(c_1)-u^P(c_2))-\gamma$ and $Q=-\gamma k_0(u^P(c_1)-u^P(c_2))$, which is the same as \eqref{20} except that $u^E(x)$ is replaced by $u^P(x)$.
	So $(0,0)$ is not a steady state.
	
	\item For the point $(0,1)$, we have $P=-(\gamma-\bar{k}\beta_1)-k_0(u^P(c_1)-u^P(c_2))$ and $Q=k_0(\gamma-\bar{k}\beta_1)(u^P(c_1)-u^P(c_2))$, which is the same as \eqref{21} except that $u^E(x)$ is replaced by $u^P(x)$.
	So $(0,1)$ is a steady state when $\bar{k}<\frac{\gamma}{\beta_1}$.
	
		\item For the state $\left(1-\frac{\gamma}{\bar{k}\beta_1},1\right)$, we have:
	\begin{equation}
	\begin{split}
	P=&-k_0(u(c_1)-u(c_2))-k_0u(c_n)\cdot\pmb{\omega} [\bar{k}\beta_1 -\gamma,\alpha],\\
	Q=&(\gamma-\bar{k}\beta_1)\{-k_0(u(c_1)-u(c_2))-(\gamma-\bar{k}\beta_1)-k_0u(c_n)\cdot\pmb{\omega} [\bar{k}\beta_1 -\gamma,\alpha]\}.
	\end{split}
	\end{equation}
	
	Since $1-\frac{\gamma}{\bar{k}\beta_1}$ is the proportion of infected individuals, we have $1-\frac{\gamma}{\bar{k}\beta_1}>0$, which is equivalent to $\bar{k}>\frac{\gamma}{\beta_1}$. (If $1-\frac{\gamma}{\bar{k}\beta_1}=0$, then the state $\left(1-\frac{\gamma}{\bar{k}\beta_1},1\right)$ becomes $(0,1)$, which we have discussed).
	To satisfy $P<0$ and $Q>0$, we first define
	\begin{equation}
	\begin{aligned}
	h_1{\buildrel \triangle \over =}&-k_0(u(c_1)-u(c_2))-k_0u(c_n)\cdot\pmb{\omega} [\bar{k}\beta_1 -\gamma,\alpha],\\ \mbox{and} \;
	h_2{\buildrel \triangle \over =}&-k_0(u(c_1)-u(c_2))-(\gamma-\bar{k}\beta_1)-k_0u(c_n)\cdot\pmb{\omega} [\bar{k}\beta_1 -\gamma,\alpha].
	\end{aligned}
	\end{equation}
	Note that $P<0$ and $Q>0$ are equivalent to $h_1<0$ and $h_2<0$.	
	It is obvious that $h_1=h_2-(\bar{k}\beta_1-\gamma)$.
	Since $(\bar{k}\beta_1-\gamma)>0$, if $h_2<0$, then $h_1<0$, and $P<0$, $Q>0$. Note that $\Phi_2=h_2$. Therefore, $\left(1-\frac{\gamma}{\bar{k}\beta_1},1\right)$ is a steady state when $\bar{k}>\frac{\gamma}{\beta_1}$ and $\Phi_2<0$.
	
	\item 
	First, as $i^{(2)}$ and $\frac{\gamma}{(1-i^{(2)})\bar{k}\beta_1}$ are the proportion of infected individuals and the proportion of individuals who choose risky behavior, by definition, we have $0\le i^{(2)} \le 1$ and $0\le \frac{\gamma}{(1-i^{(2)})\bar{k}\beta_1} \le 1$, which is equivalent to:
	\begin{equation}
	0\le i^{(2)} \le 1-\frac{\gamma}{\bar{k}\beta_1}.
	\label{eq:condi}
	\end{equation}
	
	Since $0\le i^{(2)} \le 1-\frac{\gamma}{\bar{k}\beta_1}$, it should satisfies $0<1-\frac{\gamma}{\bar{k}\beta_1}$, which means $\bar{k}>\frac{\gamma}{\beta_1}$. (If $0=1-\frac{\gamma}{\bar{k}\beta_1}$, then the state $\left(i^{(2)},\frac{\gamma}{(1-i^{(2)})\bar{k}\beta_1}\right)$ becomes $(0,1)$, which we have discussed). Then, we define $f_2(x){\buildrel \triangle \over =} k_0u^P(c_n)\cdot\pmb{\omega}(\bar{k}\beta_1 x,\alpha)-\bar{k}\beta_1 x+k_0(u^P(c_1)-u^P(c_2))$.
	By \eqref{eq:steadyPT}, $i^{(2)}$ is the solution of $f_2(x)$ so we have $f_2(i^{(2)})=0$. It is obvious that $f_2(x)$ is a monotonically decreasing function. Then $0\le i^{(2)} \le 1-\frac{\gamma}{\bar{k}\beta_1}$ is equivalent to:
	\begin{equation}
	f_2(0)\ge 0,\ \ f_2(1-\frac{\gamma}{\bar{k}\beta_1})\le 0.
	\end{equation}
	
	Note that $u^P(c_1)-u^P(c_2) >0$ and $k_0>0$. Therefore we have $f_2(0)=k_0(u^P(c_1)-u^P(c_2))> 0$. So when $\bar{k}>\frac{\gamma}{\beta_1}$ and $f_2(1-\frac{\gamma}{\bar{k}\beta_1})\le 0$, we have $0\le i^{(2)} \le 1$ and $0\le \frac{\gamma}{(1-i^{(2)})\bar{k}\beta_1} \le 1$.
	
	Then, if $\left(i^{(2)},\frac{\gamma}{(1-i^{(2)})\bar{k}\beta_1}\right)$ is steady state, it should satisfies $P<0$ and $Q>0$, so we have:
	
	\begin{equation}
	\begin{split}
	P=&-\gamma+(1-2i^{(2)})\bar{k}\beta_1x_1^{(2)}=-\bar{k}\beta_1 i^{(2)} x_1^{(2)},\\
	Q=& -\bar{k}\beta_1 i^{(2)} x_1^{(2)}(1-i^{(2)})(1-x_1^{(2)})\\
	&\cdot(k_0u^P(c_n)\pmb{\omega}'[\bar{k}\beta_1 i^{(2)},\alpha]\bar{k}\beta_1-\bar{k}\beta_1).
	\end{split}
	\end{equation}
	
	It is obvious that $P<0$.
	Since $\pmb{\omega}'[\bar{k}\beta_1 i^{(2)},\alpha] > 0$ and $k_0u^P(c_n)<0$, we have $k_0u^P(c_n)\pmb{\omega}'[\bar{k}\beta_1 i^{(2)},\alpha]\bar{k}\beta_1-\bar{k}\beta_1<0$. So $Q>0$. 
	
	Therefore, when $\bar{k}>\frac{\gamma}{\beta_1}$ and $f_2(1-\frac{\gamma}{\bar{k}\beta_1})\le 0$, $\left(i^{(2)},\frac{\gamma}{(1-i^{(2)})\bar{k}\beta_1}\right)$ is steady state. Note that $f_2(1-\frac{\gamma}{\bar{k}\beta_1})=-\Phi_2$, therefore, when 
	$\bar{k}>\frac{\gamma}{\beta_1}$ and $\Phi_2\ge 0$, $\left(i^{(2)},\frac{\gamma}{(1-i^{(2)})\bar{k}\beta_1}\right)$ is a steady state.

\end{itemize}

\section{The proof of Theorem \ref{th4}}
\label{appendix:th4}
\begin{itemize}
	\item[a.] \textbf{For Theorem 3a}: 
	
	When $\bar{k}<\frac{\gamma}{\beta_1}$, $(\bar{i}^P,\bar{x}_1^P)$ and $(\underline{i}^P,\underline{x}_1^P)$ are in Case 1 in \eqref{eq:steadyPT}. We have $(\bar{i}^P,\bar{x}_1^P)=(\underline{i}^P,\underline{x}_1^P)=(0,1)$. So all individuals, regardless of their irrationality degree, will choose the risky behavior with $\bar{x}_1^P = \underline{x}_1^P=1$
	
	\item[b.] \textbf{For Theorem 3b}:

	 When $\bar{k} >\frac{\gamma}{\beta_1}$ and $\Phi_2$ is not greater than or equal to 0 simultaneously for $\bar{\alpha}$ and $\underline{\alpha}$. 
 	Note that for the $\Phi_2$ of $\bar{\alpha}$ and $\underline{\alpha}$, there are three possibilities: 
 \begin{itemize}
     \item[1)] $\Phi_2< 0$ for both $\bar{\alpha}$ and $\underline{\alpha}$.
    \item[2)] $\Phi_2<0$ for $\bar{\alpha}$ and $\Phi_2\ge 0$ for $\underline{\alpha}$.
     \item[3)] $\Phi_2\ge 0$ for $\bar{\alpha}$ and $\Phi_2<0$ for $\underline{\alpha}$.
 \end{itemize} 
 Then we discuss these three possibilities separately.
 
 \textbf{b.1} If $1 - \frac{\gamma}{\bar{k}\beta_1} \le \frac{1}{\bar{k}\beta_1e}$, the three possibilities are: 
 \begin{itemize}
     \item[1)] If $\Phi_2< 0$ for both $\bar{\alpha}$ and $\underline{\alpha}$. Then both $(\bar{i}^P,\bar{x}_1^P)$ and $(\underline{i}^P,\underline{x}_1^P)$ are in Case 2 in \eqref{eq:steadyPT}. So all individuals, regardless of their irrationality degree, will choose the risky behavior with $\bar{x}_1^P=\underline{x}_1^P=1$.
     \item[2)] If $\Phi_2<0$ for $\bar{\alpha}$ and $\Phi_2\ge 0$ for $\underline{\alpha}$. Then $(\underline{i}^P,\underline{x}_1^P)$ are in Case 3 and $(\bar{i}^P,\bar{x}_1^P)$ are in Case 2. So all individuals with low irrationality $\bar{\alpha}$ will choose the risky behavior with $\bar{x}_1^P=1$. On the other hand, only a subset of individuals with high irrationality $\underline{\alpha}$ will opt for risky behavior with $\underline{x}_1^P < 1$.
     \item[3)] $\Phi_2\ge 0$ for $\bar{\alpha}$ and $\Phi_2<0$ for $\underline{\alpha}$ is \textbf{impossible}. In other words, if $\Phi_2<0$ for $\underline{\alpha}$, we must have $\Phi_2<0$ for $\bar{\alpha}$. Then we prove it.

      We use ${\pmb{\omega}}(x,\overline{\alpha})$ and ${\pmb{\omega}}(x,\underline{\alpha})$ to represent the weighting functions with $\overline{\alpha}$ and $\underline{\alpha}$, $\Phi_2(\bar{i}^P,\bar{x}_1^P)$ and $\Phi_2(\underline{i}^P,\underline{x}_1^P)$ to represent the $\Phi_2$ with $\overline{\alpha}$ and $\underline{\alpha}$. If $\Phi_2(\underline{i}^P,\underline{x}_1^P)<0$, we have:
	\begin{equation}
	\Phi_2(\bar{i}^P,\bar{x}_1^P)-\Phi_2(\underline{i}^P,\underline{x}_1^P)=k_0u^P(c_n)\{{\pmb{\omega}} [\bar{k}\beta_1 -\gamma,\underline{\alpha}]- {\pmb{\omega}}[\bar{k}\beta_1 -\gamma,\overline{\alpha}]\}
	\end{equation}
	
	For $\pmb{\omega}(x,\alpha)$, we have: $\frac{\partial \pmb{\omega}}{\partial \alpha}=-(-lnx)^\alpha ln(-lnx)e^{(-(-lnx)^\alpha)}$.
	And then:
	\begin{equation}
	\left\{
	\begin{aligned}
	&\frac{\partial \pmb{\omega}}{\partial \alpha}<0&& if\ \ 0<x<\frac{1}{e}\\
	&\frac{\partial \pmb{\omega}}{\partial \alpha}>0&& if\ \ \frac{1}{e}<x<1\\
	&\frac{\partial \pmb{\omega}}{\partial \alpha}=0&& if\ \ x=\frac{1}{e}.
	\end{aligned}
	\right.
	\label{eq:com_w}
	\end{equation}

	Since $1 - \frac{\gamma}{\bar{k}\beta_1} \le \frac{1}{\bar{k}\beta_1e}$, so we have $\bar{k}\beta_1 -\gamma\le \frac{1}{e}$, then based on \eqref{eq:com_w} and $u^P(c_n)<0$, we have 	$\Phi_2(\bar{i}^P,\bar{x}_1^P)-\Phi_2(\underline{i}^P,\underline{x}_1^P)\le 0$. Since $\Phi_2(\underline{i}^P,\underline{x}_1^P)<0$, we have $\Phi_2(\bar{i}^P,\bar{x}_1^P)<0$.
	So if $\Phi_2<0$ for $\underline{\alpha}$, we must have $\Phi_2<0$ for $\bar{\alpha}$.

 \end{itemize}

	\textbf{b.2} If $1 - \frac{\gamma}{\bar{k}\beta_1} \ge \frac{1}{\bar{k}\beta_1e}$, the three possibilities are: 
 \begin{itemize}
     \item[1)] If $\Phi_2< 0$ for both $\bar{\alpha}$ and $\underline{\alpha}$. Then both $(\bar{i}^P,\bar{x}_1^P)$ and $(\underline{i}^P,\underline{x}_1^P)$ are in Case 2 in \eqref{eq:steadyPT}. So all individuals, regardless of their irrationality degree, will choose the risky behavior with $\bar{x}_1^P=\underline{x}_1^P=1$.
     \item[2)] $\Phi_2< 0$ for $\bar{\alpha}$ and $\Phi_2\ge 0$ for $\underline{\alpha}$ is \textbf{impossible}. In other words, if $\Phi_2<0$ for $\bar{\alpha}$, we must have $\Phi_2<0$ for $\underline{\alpha}$. Then we prove it. If $\Phi_2(\bar{i}^P,\bar{x}_1^P)<0$, we have:
	\begin{equation}
	\Phi_2(\bar{i}^P,\bar{x}_1^P)-\Phi_2(\underline{i}^P,\underline{x}_1^P)=k_0u^P(c_n)\{{\pmb{\omega}} [\bar{k}\beta_1 -\gamma,\underline{\alpha}]- {\pmb{\omega}}[\bar{k}\beta_1 -\gamma,\overline{\alpha}]\}
	\end{equation}
	Since $1 - \frac{\gamma}{\bar{k}\beta_1} \ge \frac{1}{\bar{k}\beta_1e}$, so we have $\bar{k}\beta_1 -\gamma\ge\frac{1}{e}$, then based on \eqref{eq:com_w} and $u^P(c_n)<0$, we have $\Phi_2(\bar{i}^P,\bar{x}_1^P)-\Phi_2(\underline{i}^P,\underline{x}_1^P)\ge0$. Since $\Phi_2(\bar{i}^P,\bar{x}_1^P)<0$, we have $\Phi_2(\underline{i}^P,\underline{x}_1^P)<0$. So if $\Phi_2<0$ for $\bar{\alpha}$, we must have $\Phi_2<0$ for $\underline{\alpha}$.
	
     \item[3)] If $\Phi_2\ge 0$ for $\bar{\alpha}$ and $\Phi_2< 0$ for $\underline{\alpha}$. Then $(\underline{i}^P,\underline{x}_1^P)$ are in Case 2 and $(\bar{i}^P,\bar{x}_1^P)$ are in Case 3. So all individuals with high irrationality $\underline{\alpha}$ will choose the risky behavior with $\underline{x}_1^P=1$. On the other hand, only a subset of individuals with low irrationality $\bar{\alpha}$ will opt for risky behavior with $\bar{x}_1^P < 1$.

 \end{itemize}

	\item[c.] \textbf{For Theorem 3c}:

	For both $(\bar{i}^P,\bar{x}_1^P)$ and $(\underline{i}^P,\underline{x}_1^P)$, we have $\frac{dx_1}{dt}=0$. According to \eqref{eq:PT}, we have: 
	\begin{equation}
	\begin{split}
	\bar{x}_1^P(1-\bar{x}_1^P)[k_0u^P(c_n)\cdot\pmb{\omega}(\bar{k}\beta_1 \bar{i}^P,\overline{\alpha})-\bar{k}\beta_1 \bar{i}^P+k_0(u^P(c_1)-u^P(c_2))]=0,\\
	\underline{x}_1^P(1-\underline{x}_1^P)[k_0u^P(c_n)\cdot\pmb{\omega}(\bar{k}\beta_1 \underline{i}^P,\underline{\alpha})-\bar{k}\beta_1 \underline{i}^P
	+k_0(u^P(c_1)-u^P(c_2))]=0.
	\end{split}
	\end{equation}
	
	Since $\bar{k} >\frac{\gamma}{\beta_1}$, $\Phi_2 \geq 0$ for both $\bar{\alpha}$ and $\underline{\alpha}$, then both $(\bar{i}^P,\bar{x}_1^P)$ and $(\underline{i}^P,\underline{x}_1^P)$ are in Case 3, we have $\bar{x}_1^P\neq 0, 1$ and $\underline{x}_1^P\neq 0, 1$. We set $f_3(x)=k_0u^P(c_n)\cdot{\pmb{\omega}}(x,\overline{\alpha})-x+k_0(u^P(c_1)-u^P(c_2))$ and $f_4(x)=k_0u^P(c_n)\cdot{\pmb{\omega}}(x,\underline{\alpha})-x+k_0(u^P(c_1)-u^P(c_2))$,
	where $\pmb{\omega}(x,\overline{\alpha})$ and $\pmb{\omega}(x,\underline{\alpha})$ are the weighting functions with $\overline{\alpha}$ and $\underline{\alpha}$. Then we have:
	\begin{equation}
	\begin{split}
	f_3(\bar{k}\beta_1 \bar{i}^P)=&k_0u^P(c_n)\cdot\pmb{\omega}(\bar{k}\beta_1 \bar{i}^P,\overline{\alpha})-\bar{k}\beta_1 \bar{i}^P
	+k_0(u^P(c_1)-u^P(c_2))=0,\\
	f_4(\bar{k}\beta_1 \underline{i}^P)=&k_0u^P(c_n)\cdot\pmb{\omega}(\bar{k}\beta_1 \underline{i}^P,\underline{\alpha})-\bar{k}\beta_1 \underline{i}^P
	+k_0(u^P(c_1)-u^P(c_2))=0.
	\end{split}
	\end{equation}

	\textbf{c.1} If $0\le\bar{i}^P\le\frac{1}{\bar{k}\beta_1e}$, then we have $0\le\bar{k}\beta_1\bar{i}^P\le\frac{1}{e}$.
	According to \eqref{eq:com_w}, since $f_1(\bar{k}\beta_1 \bar{i}^P)=0$, we have $f_2(\bar{k}\beta_1 \bar{i}^P)\le0$.
	And note that $f_2(\bar{k}\beta_1 \underline{i}^P)=0$ and $f_2(x)$ is monotonically decreasing.
	So we have $\bar{i}^P\ge\underline{i}^P$.
	Then note that $\bar{x}_1^P=\frac{\gamma}{(1-\bar{i}^P)\bar{k}\beta_1}$ and $\underline{x}_1^P=\frac{\gamma}{(1-\underline{i}^P)\bar{k}\beta_1}$, so we have $\bar{x}_1^P\ge\underline{x}_1^P$.
	
	\textbf{c.2} If $\bar{i}^P\ge\frac{1}{\bar{k}\beta_1e}$, then we have $\bar{k}\beta_1\bar{i}^P\ge\frac{1}{e}$. According to \eqref{eq:com_w}, since $f_3(\bar{k}\beta_1 \bar{i}^P)=0$, we have $f_4(\bar{k}\beta_1 \bar{i}^P)\ge0$. And note that $f_4(\bar{k}\beta_1 \underline{i}^P)=0$ and $f_4(x)$ is monotonically decreasing. So we have $\bar{i}^P\le\underline{i}^P$. Then note that $\bar{x}_1^P=\frac{\gamma}{(1-\bar{i}^P)\bar{k}\beta_1}$ and $\underline{x}_1^P=\frac{\gamma}{(1-\underline{i}^P)\bar{k}\beta_1}$, so we have $\bar{x}_1^P\le\underline{x}_1^P$.

\end{itemize}
\section{The proof of Theorem \ref{th5}}
We use $(i^P,x_1^P)$ to represent the steady state after the behavior inducement.
\label{appendix:th5}
\begin{itemize}
	\item If $i^0 =0$ and $x_1^0 =1$, that is, the original steady state without behavior guidance is Case 1 in \eqref{eq:steadyPT}.
	Since both $\bar{k}$, $\gamma$, and $\beta_1$ cannot be changed through behavior inducement, no matter what the value of $\pmb{\delta}$, the steady state after behavior inducement would be $(i^P,x_1^P)=(0,1)$.
	That means, $l_1(i^P)$ and $l_2(x_1^P)$ are all constant values.
	Then we should minimize $l_3(\pmb{\delta})$.
	By definition, $l_3(\pmb{\delta})$ takes the minimum value at $\pmb{0}$.
	So the optimal solution is $\pmb{0}$.
	
	\item If $i^0 = 1-\frac{\gamma}{\bar{k}\beta_1}$ and $x_1^0 =1$, that is, the original steady state is Case 2 in \eqref{eq:steadyPT}. Since both $\bar{k}$, $\gamma$, and $\beta_1$ cannot be changed through behavior inducement, the stable state after behavior inducement $(i^P,x_1^P)$ would not be in Case 1.
	Then if $(i^P,x_1^P)$ in Case 2, then we have $(i^P,x_1^P)=(1-\frac{\gamma}{\bar{k}\beta_1},1)$, so $l_1(i^P)$ and $l_2(x_1^P)$ are all constant values.
	Then the optimal solution is $\pmb{0}$.
	If $(i^P,x_1^P)$ in Case 3, by definition, the optimal solution is $\pmb{\delta}^{(3)}$.
	In summary, the optimal solution is $\pmb{0}$ or $\pmb{\delta}^{(3)}$.
	
	\item If $0<i^0<\hat{i}$ and $0<x_1^0<1$, that is, the original steady state is Case 3 in \eqref{eq:steadyPT}. Similar to the previous analysis, $(i^P,x_1^P)$ would not be in $C_1$.
	If $(i^P,x_1^P)$ in Case 2, we have $(i^P,x_1^P)=(1-\frac{\gamma}{\bar{k}\beta_1},1)$.
	Then on the boundary of Case 3 ($\Phi_2=0$), we also have $(i,x_1)=(1-\frac{\gamma}{\bar{k}\beta_1},1)$.
	For every $(i^P,x_1^P)$ in Case 2, where the intervention vector is $\pmb{\delta}'$, We connect the initial point $\pmb{0}$ and $\pmb{\delta}'$, the line intersects the boundary of Case 2 and Case 3 at $\pmb{\delta}''$ with $(i'',x_1'')$. (By definition, it is in Case 3). Note that $(i'',x_1'')=(i^P,x_1^P)=(1-\frac{\gamma}{\bar{k}\beta_1},1)$, we have $l_1(i^P)=l_1(i'')$ and $l_2(x_1^P)=l_2(x_1'')$. Since the distance between $\pmb{\delta}''$ and $\pmb{0}$ is smaller than the distance between $\pmb{\delta}'$ and $\pmb{0}$.
	Then $l_3(\pmb{\delta}'')<l_3(\pmb{\delta}')$.
	So the loss function value of $\pmb{\delta}''$ should be smaller than $\pmb{\delta}'$. And $(i^P,x_1^P)$ cannot be the optimal solution in the whole space, which means, the optimal solution cannot be in Case 2.
	Then if $(i^P,x_1^P)$ in Case 3, by definition, the optimal solution is $\pmb{\delta}^{(3)}$.
	In summary, the optimal solution is $\pmb{\delta}^{(3)}$.
\end{itemize}

Therefore, the optimal solution in the whole space is either $\pmb{0}$ or $\pmb{\delta}^{(3)}$.

\section{The method for solving the optimization problem \eqref{eq:bound}}
\label{appendix:dL}

In \eqref{eq:bound}, for constraint \ding{175}, we use the exterior-point method as before to convert it into a penalty function. For constraint \ding{172} and \ding{173}, we use the interior-point method to convert them into penalty functions. The general form of the logarithmic barrier function of the interior-point method is $- \sum_{i}\frac{1}{t}log(-f_i(x))$, where $f_i(x)$ is the constraint. When $f_i(x)$ is close to 0, the logarithmic barrier function will have a big penalty to prevent $f_i(x)$ from getting close to 0. Then for our problem, we can convert the constraints \ding{172}, \ding{173} and \ding{175} into the following penalty function:
\begin{align}
&BF( \pmb{\delta})=-\frac{1}{t}\{\log(1-\alpha-\Delta \alpha)+\log(\alpha+\Delta \alpha)+\log(-c_n-\Delta c_n)\}+\mu \cdot P(-F_1),
\text{\ where\ } \nonumber\\ &F_1=-k_0(u^P(c_1+\Delta c_1)-u^P(c_2+\Delta c_2))-(\gamma-\bar{k}\beta_1)-k_0u^P(c_n+\Delta c_n)\cdot\pmb{\omega} [\bar{k}\beta_1 -\gamma,\alpha+\Delta \alpha]. \label{eq:BF}
\end{align}

After transforming constraints \ding{172}, \ding{173} and \ding{175} into the penalty function $BF( \pmb{\delta})$, the objective function in (\ref{eq:bound}) can be rewritten as:
\begin{equation}
\widetilde{L}=\mu [P(i^P-i_m)+P(-i^P)+ P(x_m-x_1^P)+P(x_1^P-1)]+ l_3( \pmb{\delta})+BF( \pmb{\delta}),
\end{equation}
and its gradient is:
\begin{align}
&\frac{\partial \widetilde{L}}{\partial \delta_i} = \frac{\partial l_1}{\partial i^P}\frac{\partial i^P}{\partial \delta_i}+\frac{\partial l_2}{\partial x_1^P}\frac{\partial x_1^P}{\partial i^P}\frac{\partial i^P}{\partial \delta_i}+\frac{\partial l_3}{\partial \delta_i}+\frac{\partial BF}{\partial \delta_i},
\label{eq:loss1}
\end{align}
where
$l_1(i^P)=\mu[ P(i^P-i_m)+P(-i^P)]$ and $l_2(x_1^P)=\mu[ P(x_m-x_1^P)+P(x_1^P-1)]$, $\delta_i$ is the $i$th item of vector $\pmb{\delta}$ ([$\Delta \alpha$,$\Delta c_n$,$\Delta c_1$,$\Delta c_2$]),

	\begin{algorithm}[t]
	
	\caption{The algorithm for solving the optimization problem \eqref{eq:bound}}
	
	\LinesNumbered 
	\KwIn{Initial parameter $\pmb{\delta}$, learning parameter $\eta$, $\epsilon$}
	\KwOut{$\pmb{\delta}$}
	$t=0$, set the initial $\pmb{v}$\; 
	\While{$t<iter$}
	{
		Calculate $i$ and $x_1$ by \eqref{eq:steadyPT}\;
		Calculate $\frac{\partial \widetilde{L}}{\partial \delta_i}$  based on \eqref{eq:loss1}, \eqref{45}, and \eqref{47}\;
		$v_i = \eta v_i - \epsilon \frac{\partial \widetilde{L}}{\partial \Delta \alpha}$\;
		$\delta_i = \delta_i + v_i$\;
		$t=t+1$ \;
	}
	\label{algor1}
\end{algorithm}

For the \eqref{eq:loss1}, we have:
\begin{equation}
\begin{split}
&\frac{\partial l_1}{\partial i^P}=\mu \left.\frac{dP(x)}{dx}\right|_{x=i^P-i_m}-\mu \left.\frac{dP(x)}{dx}\right|_{x=-i^P},\ \frac{\partial l_2}{\partial x_1^P}=-\mu \left.\frac{dP(x)}{dx}\right|_{x=x_m-x_1^P}+\mu \left.\frac{dP(x)}{dx}\right|_{x=x_1^P-1} ,\\
&\frac{\partial BF}{\partial \Delta \alpha}=-\frac{1}{t}\left(\frac{1}{\alpha+\Delta \alpha}+\frac{1}{\alpha+\Delta \alpha-1}\right)+\mu\cdot\left.\frac{dP(x)}{dx}\right|_{x=-F_1}\left(k_0u^P(c_n+\Delta c_n)\frac{\partial \pmb{\omega} }{\partial \alpha}|_{\bar{k}\beta_1 -\gamma,\alpha+\Delta \alpha}\right),\\
&\frac{\partial BF}{\partial \Delta c_n}=-\frac{1}{t}\cdot\frac{1}{c_n+\Delta c_n}+\mu\cdot\left.\frac{dP(x)}{dx}\right|_{x=-F_1}\left(k_0\left.\frac{d u^P(x) }{d x}\right|_{c_n+\Delta c_n} \pmb{\omega} [\bar{k}\beta_1 -\gamma,\alpha+\Delta \alpha]\right),\\
&\frac{\partial BF}{\partial \Delta c_1}=\mu\cdot\left.\frac{dP(x)}{dx}\right|_{x=-F_1}\left(k_0 \left.\frac{d u^P(x) }{d x}\right|_{c_1+\Delta c_1}\right),\ 
\frac{\partial BF}{\partial \Delta c_2}=\mu\cdot\left.\frac{dP(x)}{dx}\right|_{x=-F_1}\left(-k_0 \left.\frac{d u^P(x) }{d x}\right|_{c_2+\Delta c_2}\right).
\label{45}
\end{split}
\end{equation}
where $\frac{dP(x)}{dx}=\left\{
\begin{aligned}
&2x && if\ \ x>0\\
&0&& if\ \ x<0
\end{aligned}
\right.$, $\frac{du^P(x)}{dx}=\left\{
\begin{aligned}
&\sigma x^{\sigma-1},&& \mbox{if}\  x\ge 0,\\
&\sigma\lambda(-x)^{\sigma-1},&& \mbox{if}\  x< 0,\
\end{aligned}
\right.$, $\left.\frac{\partial \pmb{\omega} }{\partial \alpha}\right|_{x,y} = -(-lnx)^y ln(-lnx)e^{(-(-lnx)^y)}$, and  $\left.\frac{\partial \pmb{\omega} }{\partial p}\right|_{x,y} = \frac{y(-lnx)^{y-1}}{x}e^{(-(-lnx)^y)}$

According to the equality constraints, we have:
\begin{equation}
\begin{split}
&\frac{\partial i}{\partial \Delta \alpha} = \frac{-k_0u^P(c_n+\Delta c_n)\left.\frac{\partial \pmb{\omega} }{\partial \alpha}\right|_{\bar{k}\beta_1 i,\alpha+\Delta \alpha}}{\bar{k}\beta_1 k_0u^P(c_n+\Delta c_n)\left.\frac{\partial \pmb{\omega} }{\partial i}\right|_{\bar{k}\beta_1 i,\alpha+\Delta \alpha}-\bar{k}\beta_1},\\
&\frac{\partial i}{\partial \Delta c_n} = \frac{-k_0 \left.\frac{d u^P(x) }{d x}\right|_{c_n+\Delta c_n}\pmb{\omega}[\bar{k}\beta_1 i,\alpha+\Delta \alpha]}{\bar{k}\beta_1 k_0u^P(c_n+\Delta c_n)\left.\frac{\partial \pmb{\omega} }{\partial i}\right|_{\bar{k}\beta_1 i,\alpha+\Delta \alpha}-\bar{k}\beta_1},\\
&\frac{\partial i}{\partial \Delta c_1} = \frac{-k_0\left.\frac{d u^P(x) }{d x}\right|_{c_1+\Delta c_1}}{\bar{k}\beta_1 k_0u^P(c_n+\Delta c_n)\left.\frac{\partial \pmb{\omega} }{\partial i}\right|_{\bar{k}\beta_1 i,\alpha+\Delta \alpha}-\bar{k}\beta_1},\\
&\frac{\partial i}{\partial \Delta c_2} = \frac{k_0\left.\frac{d u^P(x) }{d x}\right|_{c_2+\Delta c_2}}{\bar{k}\beta_1 k_0u^P(c_n+\Delta c_n)\left.\frac{\partial \pmb{\omega} }{\partial i}\right|_{\bar{k}\beta_1 i,\alpha+\Delta \alpha}-\bar{k}\beta_1},\\
&\frac{\partial x_1}{\partial i} = \frac{\gamma}{(1-i)^2\bar{k}\beta_1}
\label{47}
\end{split}
\end{equation}

Since $l_3( \pmb{\delta})$ are known and derivable,
then $\frac{\partial l_3}{\partial \pmb{\delta}}$ are known.
Then we can calculate the gradient $\frac{\partial \widetilde{L}}{\partial \delta_i}$ based on \eqref{eq:loss1}, \eqref{45}, and \eqref{47}. 

Since we cannot guarantee the convexity of the objective function, to improve the convergence speed and jump out of the local optimum, we optimize the problem of \eqref{eq:bound} with the Momentum Gradient Descent method\cite{polyak1964some}. 

In summary, we can solve the optimization problem \eqref{eq:bound} by Algorithm 1.

\section{The method for solving the optimization problem \eqref{eq:unbound}}
\label{appendix:unbound}
Similar to the optimization problem \eqref{eq:bound}, the objective function in \eqref{eq:unbound} can be rewritten as:
\begin{equation}
\widetilde{L}_2=(i^P-i_m)^2+(x_1^P-x_m)^2+ l_3( \pmb{\delta})+BF( \pmb{\delta}),
\end{equation}
where $BF( \pmb{\delta})$ is defined in \eqref{eq:BF}, and its gradient is:
\begin{align}
&\frac{\partial \widetilde{L}_2}{\partial \delta_i} = \frac{\partial l_1}{\partial i^P}\frac{\partial i^P}{\partial \delta_i}+\frac{\partial l_2}{\partial x_1^P}\frac{\partial x_1^P}{\partial i^P}\frac{\partial i^P}{\partial \delta_i}+\frac{\partial l_3}{\partial \delta_i}+\frac{\partial BF}{\partial \delta_i},\label{gradient2}\\
\text{where }& l_1(i^P)=(i^P-i_m)^2 \text{ and } l_2(x_1^P)=(x_1^P-x_m)^2.\nonumber
\end{align}

For the \eqref{gradient2}, we have:
\begin{equation}
\begin{split}
&\frac{\partial l_1}{\partial i^P}=2(i^P-i_m),\ \frac{\partial l_2}{\partial x_1^P}=2(x_1^P-x_m) ,\\
&\frac{\partial BF}{\partial \Delta \alpha}=-\frac{1}{t}\left(\frac{1}{\alpha+\Delta \alpha}+\frac{1}{\alpha+\Delta \alpha-1}\right)+\mu\cdot\left.\frac{dP(x)}{dx}\right|_{x=-F_1}\left(k_0u^P(c_n+\Delta c_n)\frac{\partial \pmb{\omega} }{\partial \alpha}|_{\bar{k}\beta_1 -\gamma,\alpha+\Delta \alpha}\right),\\
&\frac{\partial BF}{\partial \Delta c_n}=-\frac{1}{t}\cdot\frac{1}{c_n+\Delta c_n}+\mu\cdot\left.\frac{dP(x)}{dx}\right|_{x=-F_1}\left(k_0\left.\frac{d u^P(x) }{d x}\right|_{c_n+\Delta c_n} \pmb{\omega} [\bar{k}\beta_1 -\gamma,\alpha+\Delta \alpha]\right),\\
&\frac{\partial BF}{\partial \Delta c_1}=\mu\cdot\left.\frac{dP(x)}{dx}\right|_{x=-F_1}\left(k_0 \left.\frac{d u^P(x) }{d x}\right|_{c_1+\Delta c_1}\right),\ 
\frac{\partial BF}{\partial \Delta c_2}=\mu\cdot\left.\frac{dP(x)}{dx}\right|_{x=-F_1}\left(-k_0 \left.\frac{d u^P(x) }{d x}\right|_{c_2+\Delta c_2}\right).
\label{eq:H3}
\end{split}
\end{equation}

Then we can get the $\frac{\partial \widetilde{L}_2}{\partial \delta_i}$ based on \eqref{gradient2}, \eqref{eq:H3} and \eqref{47}. We can optimize the problem of \eqref{eq:unbound} with the Momentum Gradient Descent method in Algorithm 2.
\begin{algorithm}[t]
	
	\caption{The algorithm for solving the optimization problem \eqref{eq:unbound}}
	
	\LinesNumbered 
	\KwIn{Initial parameter $\pmb{\delta}$, learning parameter $\eta$, $\epsilon$}
	\KwOut{$\pmb{\delta}$}
	$t=0$, set the initial $\pmb{v}$\; 
	\While{$t<iter$}
	{
		Calculate $i$ and $x_1$ by \eqref{eq:steadyPT}\;
		Calculate $\frac{\partial \widetilde{L}_2}{\partial \delta_i}$ based on \eqref{gradient2}, \eqref{eq:H3} and \eqref{47} \;
		$v_i = \eta v_i - \epsilon \frac{\partial \widetilde{L}_2}{\partial \Delta \alpha}$\;
		$\delta_i = \delta_i + v_i$\;
		$t=t+1$ \;
	}
	\label{algor1}
\end{algorithm}
\section{Analysis of the 3-behavior model}
\label{appendix:3behavior}

For the model with three behaviors, theoretical analysis alone becomes challenging, prompting us to resort to simulations and numerical analysis. In our study, we employ networks with 500 nodes. The physical contact network has a fixed degree of 10, while the information network has a degree of 20. We use \label{utilityequation} as the utility function, with $\sigma=0.65$ and $\lambda=1$. Additionally, we set $\gamma=0.03$ and $c_n=-20$.

\begin{figure}[t]
	
	\centering
	\subfigure[]{
		\begin{minipage}[t]{0.5\linewidth}
			\centering
			\includegraphics[width=1\linewidth]{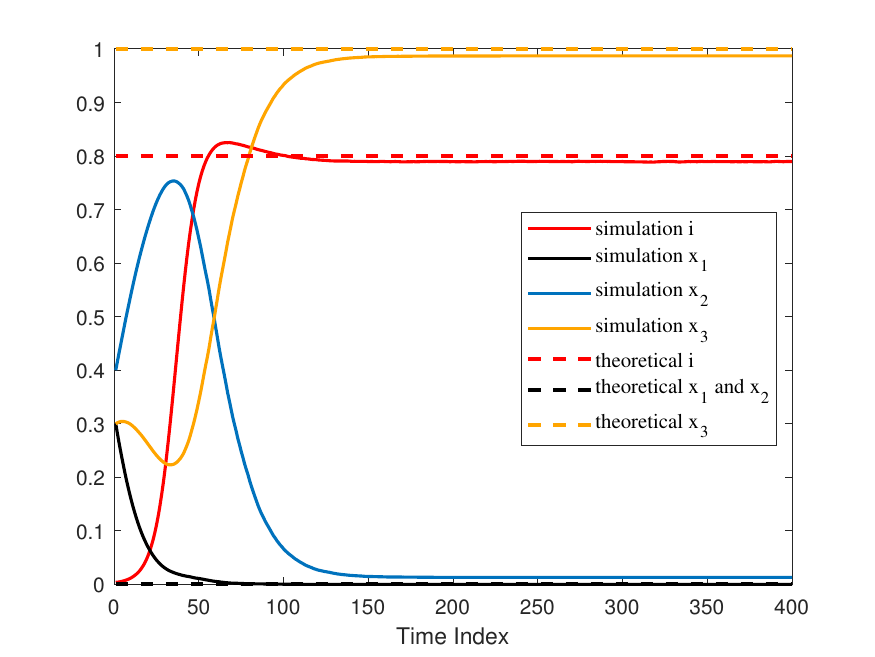}
			\label{fig:3behavior1}
		\end{minipage}%
	}%
	\subfigure[]{
		\begin{minipage}[t]{0.5\linewidth}
			\centering
			\includegraphics[width=1\linewidth]{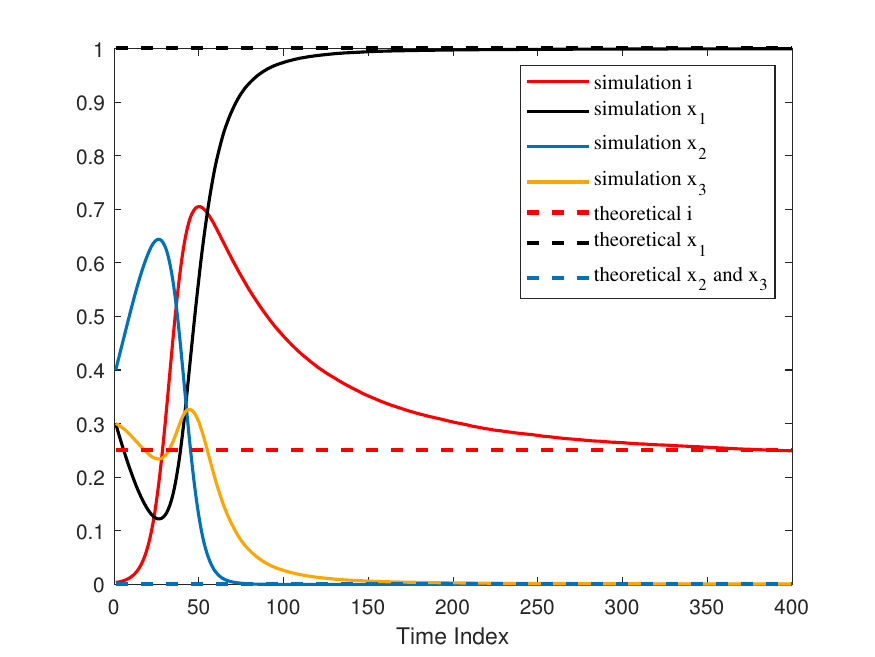}
			\label{fig:3behavior2}
		\end{minipage}%
	}%
	
	\centering
	\caption{The simulation and theoretical results of the 3-behavior model.
	}
	\label{fig:3behavior}
\end{figure}

For the three behaviors $a_1$, $a_2$, and $a_3$, we set $\{c_1,c_2,c_3\}=\{-4.5,-2,-1\}$ and $\{\beta_1,\beta_2,\beta_3\}=\{0.005,0.015,0.025\}$. The simulation results, along with the corresponding theoretical outcomes, are presented in Fig. \ref{fig:3behavior1}. Subsequently, we alter the values to $\{c_1,c_2,c_3\}=\{-3,-2,-1\}$ and $\{\beta_1,\beta_2,\beta_3\}=\{0.004,0.015,0.031\}$, and the respective results are displayed in Fig. \ref{fig:3behavior2}. Notably, the excellent agreement between the theoretical and simulation results validates the applicability of our model in the context of multi-behavior scenarios.

\section{Irrationality makes individuals more risk-seeking}

\label{appendix:radical}
\begin{figure}[htbp]
	
	\centering
	\subfigure[]{
		\begin{minipage}[t]{0.5\linewidth}
			\centering
			\includegraphics[width=0.9\linewidth]{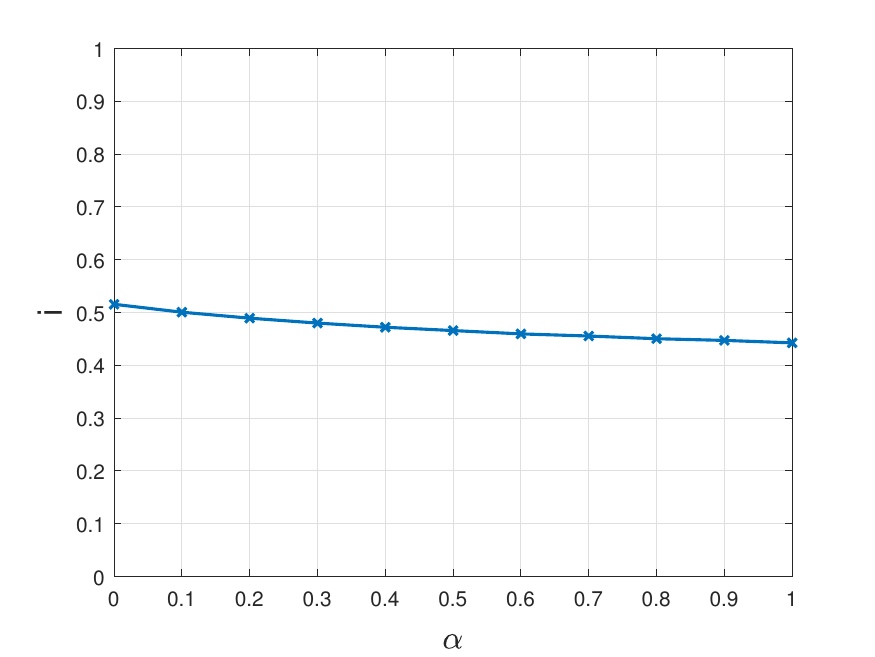}
			\label{Alpha_high_i}
		\end{minipage}%
	}%
	\subfigure[]{
		\begin{minipage}[t]{0.5\linewidth}
			\centering
			\includegraphics[width=0.9\linewidth]{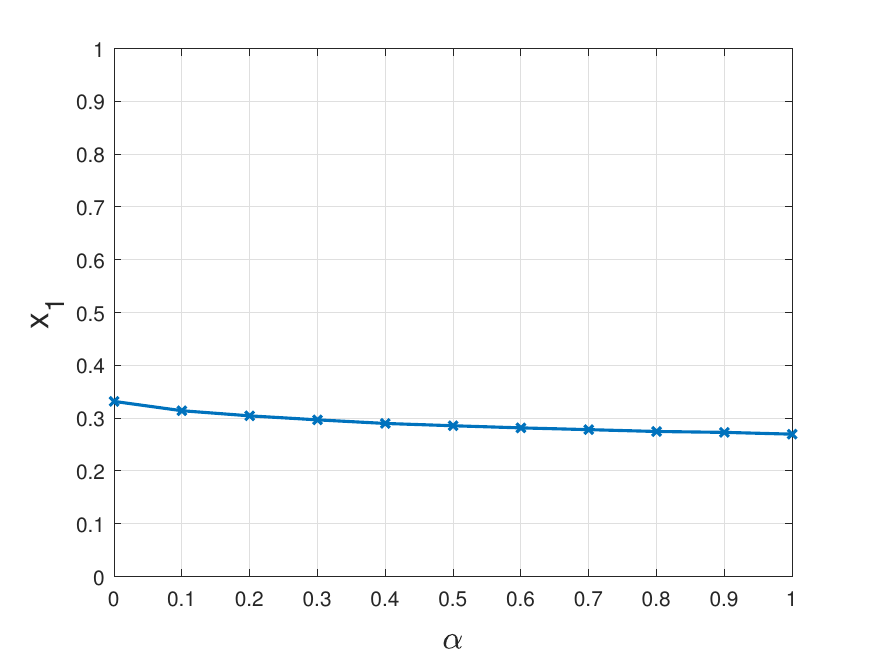}
			\label{Alpha_high_x}
		\end{minipage}%
	}%
	
	\centering
	\caption{The simulation results of $i$ and $x_1$ when changing $\alpha$.
	}
	\label{Alpha_high}
\end{figure}

In the experiments, drive individuals to be risk-seeking only in rare cases. We run simulations on regular networks with 500 nodes. The physical contact network has a fixed degree of 10, while the information network has a degree of 20. We set $c_1=0$, $c_2=-1$, $c_n=-1.1$, $\beta_1=0.1$ and $\gamma=0.12$. The simulation results are shown in Fig. \ref{Alpha_high}. All three curves belong to Case 3, and they have $\frac{1}{\bar{k} \beta_1 e}<1$. It can be seen that in this situation, when the degree of irrationality increases, the proportion of risky behavior increases. Through experiments, it is found that this situation only occurs when the loss of disease is very low, almost the same as the gain of conservative behavior ($c_2=-1$ and $c_n=-1.1$), and the infection rate and recovery rate are high. And in this situation, although the irrational factor makes the individual risk-seeking, this effect is not obvious.

\section{The analysis of the situations where irrationality promotes individuals to become risk-seeking}
\label{appendix:radicalwhy}
Through experiments, we find that the situation where irrationality makes individuals risk-seeking only occurs when the infection rate is high, and the loss of disease is extremely low. In this section, we analyze this phenomenon.

According to Theorem \ref{th4}c, if $\bar{i}^P\ge\frac{1}{\bar{k}\beta_1 e}$ and $\bar{\alpha}>\underline{\alpha}$, we have $\bar{i}^P\le \underline{i}^P,\bar{x}_1^P\le \underline{x}_1^P$, which is the situation where irrationality makes individuals risk-seeking. For simplicity, we set $\bar{\alpha}=1$, then according to Theorem \ref{th2} and Theorem \ref{th4}, we have:

\begin{equation}
\begin{split}
& \bar{i}^P =\frac{k_0(u^{P}(c_2)-u^{P}(c_1))}{(k_0u^{P}(c_n)-1)\bar{k}\beta_1}\ge \frac{1}{\bar{k}\beta_1 e},\\
&-k_0(u^{P}(c_1)-u^{P}(c_2))+(\gamma-\bar{k}\beta_1)(k_0u^{P}(c_n)-1) \ge 0.
\end{split}
\label{eq:54}
\end{equation}

Simplify the \eqref{eq:54} and we have:
\begin{equation}
\begin{aligned}
& \bar{k}\beta_1\ge\frac{1}{e}+\gamma,&&\text{\ding{172}}\\
&\frac{u^{P}(c_n)}{u^{P}(c_2)-u^{P}(c_1)} < e+\frac{1}{k_0(u^{P}(c_2)-u^{P}(c_1))}.&&\text{\ding{173}}
\end{aligned}
\label{eq:55}
\end{equation}

Based on \eqref{eq:55}-\ding{172}, $\beta_1$ or $\bar{k}$ need to be very high (compared to $\beta_1 \ll 1$). In short, for an individual, the probability of being infected is very high. And based on \eqref{eq:55}-\ding{173}, since $u^{E}(c_2)-u^{E}(c_1) < 0$ and $k_0<1$, $\frac{u^{P}(c_n)}{u^{P}(c_2)-u^{P}(c_1)}$ is lower than a bound that is smaller than $e$, which means, compared with the payoff gap between the two behaviors, the loss of being infected is small.

Therefore, the situation where irrationality makes individuals more radical only occurs when the infection rate is high, and the loss of disease is extremely low.

\end{document}